\pdfoutput=1 
\documentclass[final,1p,times]{elsarticle}
\usepackage{geometry}
\newgeometry{vmargin={20mm}, hmargin={20mm,15mm}}  
\graphicspath{{Fig/}}
\usepackage{graphics}
\usepackage{graphicx}
\usepackage{epstopdf} 
\usepackage{floatrow}
\usepackage{amssymb}
\usepackage{natbib}
\usepackage{amsthm}
\usepackage{amsmath}
\usepackage{multirow}
\usepackage{lineno}
\usepackage{placeins}
\usepackage[export]{adjustbox}
\usepackage{xcolor}
\usepackage{color}
\usepackage[colorlinks=true]{hyperref}
\usepackage{hyperref}
\usepackage{url}
\usepackage{tikz}

\usepackage[utf8]{inputenc}

\RequirePackage{snapshot}	
\setcitestyle{authoryear,open={(},close={)}}

\begin{document}

\immediate\write18{dir}
\begin{frontmatter}

\title{Numerical investigation of the nonlinear interaction between the sinusoidal motion-induced and gust-induced forces acting on bridge decks}

%\author{Samuel Tesfaye$^{*}$, Igor Kavrakov,  Guido Morgenthal}

\author[]{Samuel Tesfaye\corref{cor1}}
\ead{samuel.tesfaye.gebretensaye@uni-weimar.de}
\cortext[cor1]{Corresponding author, Tel.: +49 3643 584426}
\author[]{Igor Kavrakov}
\author[]{Guido Morgenthal}

\address{Bauhaus University Weimar, Institute of Structural Engineering, Chair of Modelling and Simulation of Structures, \\Marienstr. 7A, 99423 Weimar, Germany }

%\address{Bauhaus University, Weimar}

\begin{abstract}

With the increasing spans and complex deck shapes, aerodynamic nonlinearity becomes a crucial concern in the design of long-span bridges. This paper investigates the nonlinear interaction between the gust-induced and motion-induced forces acting on bridge decks using the Vortex Particle Method (VPM) as a Computational Fluid Dynamics (CFD) method. This nonlinear interaction is complex and intractable by the conventional linear semi-analytical models that employ the superposition principle. To excite such interaction, a sinusoidally oscillating bridge deck is subjected to sinusoidal vertical gusts. Two distinct aspects are studied: The influence of large-scale sinusoidal vertical gusts on the shear layer of a moving body and the nonlinear dependence of the aerodynamic forces on the effective angle of attack. For the latter, the resultant aerodynamic forces based on a linear semi-analytical and a CFD model are compared for the same effective angle of attack due to sinusoidal gust and motion. The methodology is first employed to verify the linear behavior of a flat plate and then study the nonlinear behavior of two bridge decks.The results show that the linear superposition principle holds for streamlined bridge decks with minimal shear layer instability. However, this principle may not be valid for bluff bridge decks due to strong separation vortices that dictate the shear layer, which induces nonlinearity in the aerodynamic forces manifested through amplitude difference in the first harmonic and emergence of higher-order harmonics. The outcome of this study aims to provide a deeper understanding of the complex nonlinear fluid-structure interaction occurring for bluff bodies subjected to motion and free-stream gusts.

\end{abstract}

\end{frontmatter}
%
%\linenumbers
\section{Introduction}
\label{Sec1_Intro}
Wind-induced vibrations are commonly the leading design concern for long-span bridges; thus, there is a need for reliable and efficient prediction models.  Along with wind tunnel testing and computational fluid dynamics (CFD), semi-analytical models are an integral part of the wind load analysis framework. Ideally, wind-induced vibration problems should be modeled synergistically in turbulent atmospheric conditions, but the dynamic fluid-structure interaction is ordinarily approximated into sub-groups based on the linearity premise. To this end, aerodynamic forces on a bridge deck are commonly represented as a linear superposition of the static force (induced by the mean wind flow), gust-induced force (buffeting), and motion-induced force (self-excited). 

The linear unsteady (LU) semi-analytical model rests on the foundation of the airfoil theory, wherein the unsteady gust-induced \citep{Davenport1962} and motion-induced \citep{Scanlan1971a} forces are superimposed for moderately streamlined bridge decks under the small amplitude threshold, the aerodynamic forces are considered linear \citep{Scanlan1993}. The unsteady effects that originate from the linear fluid memory in the motion-induced forces and the uncorrelated chord-wise velocity distribution are accounted by frequency-dependent aerodynamic coefficients in terms of aerodynamic derivatives and aerodynamic admittance, respectively \citep{Tubino2005}. Employing the superposition principle, these aerodynamic coefficients are determined separately using wind tunnel experiments (e.g., \citep{Iwamoto1995,Chen2005,Larose1999}) or CFD simulations (e.g., \citep {Larsen1998,Kavrakov2019}).

For conventional bridge decks, the LU model has proven its utility for a wide range of global metrics with acceptable accuracy \citep{Wu2013a,Kavrakov2017}. It is now well-established that aerodynamic coefficients are a function of the input amplitude, wind angle of attack, and in some cases mode of vibration \citep {Sankaran1992,Noda2003,Matsumoto1993a}. Nevertheless, there is no concise finding on the effects of free-stream turbulence on motion-induced force or vice-versa.  Generally, the LU model does not account for the nonlinear interaction between free-stream gust and a moving body, which manifests itself through the nonlinear aerodynamic forces and strong shear layer instability. In this study, shear layer instability describes the breaking of the surface boundary layer to fully turbulent, leading to unsteady flow separation and reattachment.

To date, different nonlinear aerodynamic models have been put forward. Models based on nonlinear wake-oscillator \citep{
Tamura2020}  were used to predict the combined vortex-induced and galloping vibration of bluff cylinders in laminar flow \citep{
Corless1988, Mannini2018, Marra2011}. Recently, \citep{
Mannini2020} introduced a quasi-steady free stream-turbulence expression in the wake-oscillator model to analyze unsteady galloping of a rectangular cylinder. Models based on the Volterra series employed in the study of vortex-induced and motion-induced vibration problems \citep{Carassale2010, Nprstek2007, Cheng2017, Wu2013}.  One prospect of the Volterra nonlinear convolution scheme is its capability to model the higher-order fluid memory that is apparent in nonlinear aerodynamic systems.  Nonlinear mathematical formulations, the Taylor and Van der Pol type equations, also applied in motion-induced limit cycle oscillation studies \citep{Zhu2020, Nprstek2007, Nprstek2020, Bolotin2002a}. Following the latest advancements in machine learning, data-driven techniques have  surfaced in modeling the aerodynamics of bluff bodies \citep{Kavrakov2022, Wu2011, Abbas2020}. Various proposed methods attempt to account for the effective angle of attack, including the contribution from free stream turbulence and motion, in the time domain, among which are included the rheological \citep{Diana2008} and hybrid \citep{Chen2003a, Diana2013} models. 
     
%This study investigates the nonlinear interaction between the unsteady gust-induced and motion-induced forces using the Vortex Particle Method (VPM) as a CFD scheme.

Herein, we simulated the wind-induced vibration problem in a turbulent environment in a fully coupled manner using the Vortex Particle Method (VPM) as a CFD scheme and investigate the nonlinearity that originated from the interaction of  the unsteady motion- and gust-induced forces. \color{black} A sinusoidally oscillating bridge deck is subjected to free-stream sinusoidal gusts to excite a nonlinear interaction between the gust-induced and motion-induced forces at large angles of attack. The interaction is examined by comparing the time-dependent forces of the CFD model with their linear counterpart from the LU model with CFD-based aerodynamic coefficients. To this end, the influence of the free-stream turbulence on the shear layer is discussed in an attempt to explain the origin of the aerodynamic nonlinearity in the forces based on peculiar flow features. The paper is organized as follows: Section \ref{Sec2_Litrature} briefly discusses the fundamental aspects of the influence of free-stream turbulence on the aerodynamic forces of a moving body reported in the literature. Section \ref{Sec2_Method} describes the methodology used to study the interaction, and it revisits the CFD and LU models, including previously-developed comparison metrics for the force time-histories. Verification of the numerical procedure for linear aerodynamic of a flat plate is presented in Sec. \ref{Sec3_Verif}. In Sec. \ref{Sec4_Appl}, the methodology is applied on two generic bridge decks to excite the nonlinear interaction and the results discussed in depth. Finally, Sec. \ref{Sec7_Conlusion} concludes the study. 

\section{Influence of Free-stream Turbulence on the Aerodynamic Forces of a Moving Body}
\label{Sec2_Litrature}
\begin{figure}[b!]
\begin{minipage}[h!]{0.49\textwidth}
\centering
\includegraphics[trim=0.11cm 0cm 11.4cm 0.1cm,clip,width=0.8\columnwidth]{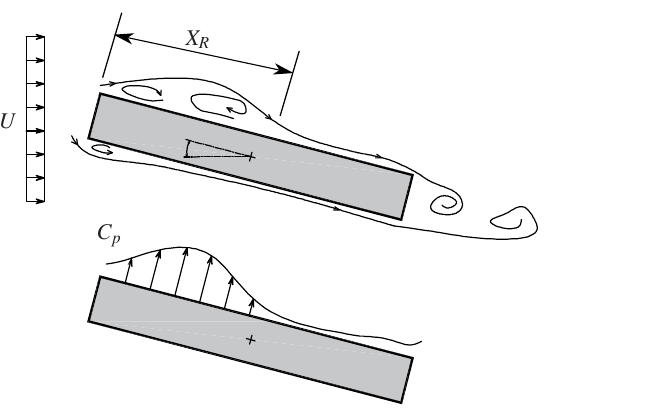}
\end{minipage}
%\hfill
\begin{minipage}[h!]{0.49\textwidth}
\centering
\includegraphics[trim=11.4cm 0cm 0.11cm 0.1cm,clip,width=0.8\columnwidth]{Fig1}
\end{minipage}
\caption{ Schematics of the instantaneous shear layer and the corresponding instantaneous top-chord motion-induced pressure distribution. For a forced sinusoidal pitch motion in laminar (left) and turbulent (right) flow cases.  }
\label{sketch_pr}
\end{figure}

The nonlinear behavior in bluff bridge decks originates from unsteady flow separation and reattachment and the ensuing wake dynamics \citep{Wu2013b}. Instantaneous pressure records inside separation bubbles revealed transient amplitude outbursts on a quasi-periodic background fluctuation \citep{Cherry1984}.  The low-frequency content is roughly in agreement with the average shedding frequency of large-scale eddies.  The amplitude outburst corresponds to the convection of successive small-scale vortex cores close to the surface. Numerous fundamental studies on static rectangular prism show that; small-scale turbulence (relative to body size) shortens the mean separation bubble up to 40\% compared to the laminar flow, increases the peak pressure amplitude, and shifts the maximum mean pressure location towards the reattachment point \citep{Hillier1981,Kiya1983,Saathoff1989}.  These effects were found to increase along with turbulence intensity. Large-scale turbulence with an integral scale up to $7.8D$ (where $D$ is the frontal depth) manifests itself on the mean pressure distribution \citep{HaanJr1998}.  Further increases in scale rapidly rise towards the smooth flow value. Similar behavior observed for a harmonically rotating rectangular prism \citep {HaanJr2009}, the reattachment point $X_R$ shifts to the leading edge, as shown in Fig.~\ref {sketch_pr}.

\FloatBarrier
The incident turbulence modifies both the top-chord instantaneous motion-induced pressure $C_P$ and the phase distribution between the body and the induced force. The effect on the phase is more pronounced for the change of intensity than the integral scale of the incoming turbulence. Nonetheless, the peak pressure locale aligns with the rapidly increasing phase region. Thus, in addition to the localized aerodynamic load effects, the separation region significantly impacts the aeroelastic stability behavior. Studying the aerodynamic derivatives, formulated from the chord-wise self-excited pressure amplitude and phase, reveals that turbulence has a stabilizing effect and decreases the motion-induced lift force. In the case of buffeting forces, the oscillation of the section increases the upstream gust-induced pressure, resulting in an average 10\% increase of the buffeting forces compared to the corresponding stationary system \citep {HaanJr2009}.

Although the interaction is more likely to result in a nonlinear behavior, such as multiple frequency excitation and harmonic distortion, past research on bridge decks focused on the influence of incident turbulence on aerodynamic derivatives. For bluff box girders, turbulence shows a stabilizing effect by shifting the reduced velocity at which the torsional damping ratio becomes positive \citep {Lin2005}.  This behavior tends to increase monotonically with the turbulence intensity. On the contrary, \cite{Sarkar1994}  concluded turbulence on streamlined bridge decks does not have an appreciable effect on the aerodynamic derivatives. However, they pointed out that the turbulence effect is present as noise in the measured signal, and the conclusion is sensitive to the identification procedure. Another aspect to be considered is the turbulence structure of the approaching flow. Wind tunnel study on the Golden Gate Bridge deck shows large-scale vertical gusting generated using flapping wings have a destabilizing effect \citep{Huston1986}. This is different from the trend observed for $\mathrm{H}$ and truss sections amidst grid-generated turbulence  \citep{Scanlan1978,Lam2017} and requires attention, as long-span bridges could experience turbulence with length scales as large as ten times their deck width. 

Another concern worthy of investigation is the variation of the aerodynamic derivatives to the effective angle of attack. Experimental findings show even for small amplitudes of structural motion, free-stream turbulence may increase the effective angle enough to result in significant variation of the aerodynamic derivatives. This behavior depends on the particular way or sensitivity in which the aerodynamic derivatives change with the angle of attack. Box-shaped \citep{Diana1993}, due to their unsymmetric geometry, and multi-box  \citep{Diana2013,Zasso2019} bridge decks are some of the sections that exhibit nonlinear dependence on the aerodynamic coefﬁcients with the effective angle of  attack. 

%Various proposed methods attempt to account for the effective angle of attack in the time domain, among which are included the artificial neural network \citep {Wu2011}; rheological \citep {Diana2008} and hybrid \citep{Chen2003a,Diana2013} models.

In light of the above discussion, it is evident that investigation of the aerodynamic derivatives in turbulent flow would not comprehensively address the interaction of the gust and motion-induced forces. More importantly, the experimental studies only address the influence of free-stream turbulence on the linear term of the fluid memory \citep{Chen2005}. Thus, this study offers both quantitative and qualitative assessments of the linear superposition principle.

\section{Methodology}
\label{Sec2_Method}
This section describes the components of the presented framework used to study the interaction between the gust-induced and motion-induced forces acting on a bluff body. These components include: (i) CFD model, used to simulate nonlinear aerodynamics; (ii) LU semi-analytical model, used to simulate linear aerodynamics based on CFD aerodynamic coefficients; and (iii) comparison metrics for time-histories, used to quantitatively assess the discrepancies between the linear and nonlinear aerodynamic forces.
\begin{figure}[!h]
\centering
\includegraphics[trim=0cm 0cm 0cm 0cm,clip,width=1\columnwidth]{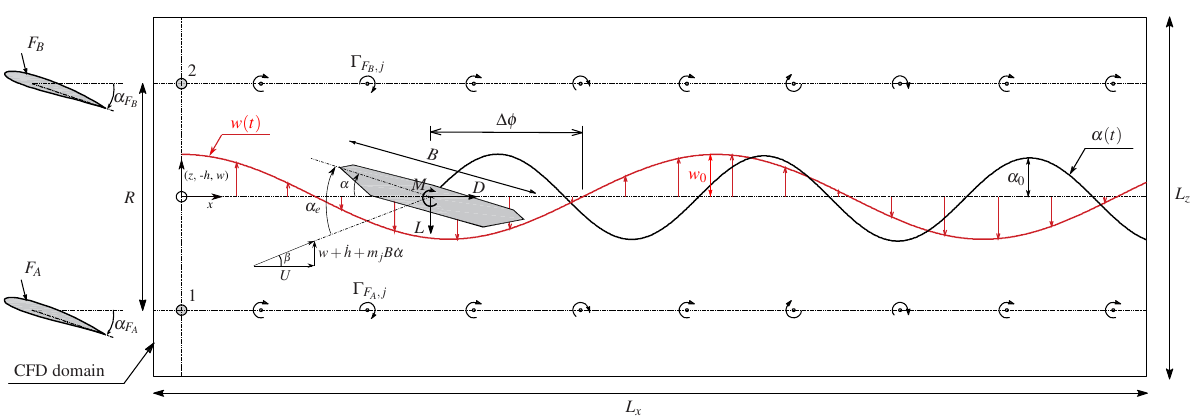} 
\caption{ Setup: A sinusoidally oscillating bluff body under a free-stream sinusoidal vertical gust.}
\label{Fig2}
\end{figure}
\FloatBarrier

\subsection{Setup}
The setup is depicted in Fig.~\ref{Fig2}: A body is performing forced sinusoidal oscillations and is concurrently subjected to free-stream sinusoidal vertical gusts. The oscillation is either in the heave or pitch degree of freedom (DOF). Additionally, two cases, each with a distinct phase-shift between the input motion and incoming gust, are considered to enrich the interaction. The interaction is studied from two aspects. First, the shear layer is qualitatively investigated by looking at the pressure and velocity fields based on a setup with a single excitation (gust or motion) and a setup with combined excitation. Second, the aerodynamic forces are studied based on the effective angle of attack comprised of gust and motion. 

The single frequency harmonic gust allows to regulate the effective angle of attack and determine the complex form of aerodynamic admittance. The instantaneous effective angle of attack $\alpha_e=\alpha_e(t)$, time $t$, according to the quasi-stead formulation is given as \citep{Chen2002}:
\begin{equation}
 \label{effctangle}
\begin{split}
\alpha_e= \alpha_s + \alpha + \beta = \alpha_s + \alpha + \arctan \left( \; \frac{w + \dot{h} + m_j B \dot{\alpha}} {U} \; \right),
\end{split}
\end{equation}
which for small angles of attack and single frequency $f$ excitation of gust and motion can be rewritten in the heave and pitch DOF, respectively, as:
\begin{equation}
 \label{effectAngl2}
\begin{aligned}
\alpha_e&= \alpha_s + \frac{\dot{h}_0}{U}  \; cos(2 \pi f t ) + \frac{w_0}{U} \;  cos(2 \pi f t + \Delta \phi_{w, \dot{h}} ), \\
\alpha_e&=\alpha_s + \alpha_{tot} \; cos(2 \pi f t ) + \frac{w_0}{U} \; cos(2 \pi f t  + \Delta \phi_{w, \alpha}), 
\; \;  \; \; 
%\alpha_{tot} =  \left( \alpha_0  +  \;  \frac{m_j B \dot{\alpha}_0} {U} \; \right) 
\alpha_{tot} =   \alpha_0  +  \;  \frac{m_j B \dot{\alpha}_0} {U}, \;  
\end{aligned}
\end{equation}

where $B$ is the bridge deck width, $U$ is the mean wind velocity, $\alpha_s$ is the static angle of attack ($\alpha_s$=0 in this study). $\alpha$=$\alpha(t)$  is the time varying forced pitch oscillation with amplitude $\alpha_0$. $\dot{h}_0$ and $\dot{\alpha}_0$ are respectively the heave and pitch oscillation velocity amplitudes, and $w_0$ is the gust velocity amplitude as shown in Fig.~2. $\alpha_{tot}$ is the total pitch angle accounting for the pitch velocity taken at a specific point on the deck defined by the parameter $m_j$. The phase shifts between the gust and motion, heave or pitch, are given as $\Delta \phi_{w,\dot{h}}$ and $\Delta \phi_{w,\alpha}$, respectively.   For sinusoidal pitch oscillation with amplitude $\alpha_0$, the velocity amplitude $\dot{\alpha}_0$=$2 \pi f \alpha_0$ and  $\alpha_{tot}$ in Eq.~(\ref{effectAngl2}) would read,

\begin{equation}
 \label{eff_alpha_0}
\begin{aligned} 
\alpha_{tot} = \alpha_0 \left(  1+\frac{m_j 2 \pi}{V_r} \right),
\end{aligned}
\end{equation}

where $V_r= U/(Bf)$ is the reduced velocity for input signal frequency $f$.

The parameter $m_j$ for $j \in \{ L, M\}$, where $L$ is the lift and $M$ is the moment force, defines the location of the normal ﬂuid velocity component (downwash) used to calculate the aerodynamic forces based on the quasi-steady aerodynamic model. In the case of the analytical quasi-steady model of a flat plate, the aerodynamic center is at the third quarter-point $m_j$=0.25  \cite{Fung2008}. In the case of bridge decks, the aerodynamic forces depend on multiple downwash points due to the flow separation and reattachment. Thus, the parameter $m_j$ defines an equivalent aerodynamic steady-state \cite{VanOudheusden1995, Kavrakov2016} and is often obtained based on the aerodynamic derivatives \cite{Diana1993}, generally having values between -0.5 and 0.5 \cite{Wu2013a}. Herein, the downwash location is only indicative when specifying the total angle of attack $\alpha_{tot}$ as an input, i.e. the quasi-steady model is not employed; thus, the aerodynamic center is set constant to $m_L$=$m_M$=0.25 for simplicity. Moreover, the contribution of the term related to $m_j$ to the total angle $\alpha_{tot}$ is generally small for the selected $V_r$ range based on Eq.~(3) for both bridge decks. However, it is noted that using the quasi-steady aerodynamic model with a positive value for $m_M$ for streamlined decks may lead to premature flutter during aeroelastic analysis \cite{Wu2013a}.

\color {black}

The phase and amplitude regulate the amplitude of the effective angle of attack. Table \ref{T1} depict the combination of input parameters used in this study. A total of 14 cases are studied, wherein the first 6 (I-VI) are with a single type of excitation with two different effective angle amplitudes of 1.5 and 4 degrees. Suppose the CFD-based aerodynamic coefficients (aerodynamic derivatives and aerodynamic admittance) are similar for two different amplitudes. In that case, it can be reasoned that the linear hypothesis used in the LU model is retained. This is important, as the goal is to keep the separate contribution of the motion- and gust-induced forces in the linear range while studying their nonlinear interaction when these two contributions are co-acting. Cases VII-XIV are designed to study this interaction, and they represent a superposition of the individual excitation cases (I-VI). The total effective angle ranges from 2 to 8 degrees, considering the two cases with distinct phase-shift between the motion and gust sinusoids. The idea behind selecting a wide range of cases with different amplitudes is to systematically study the interaction of sinusoidal gust and motion with identical frequency.    
\begin{table}[h]
\centering
\renewcommand{\arraystretch}{1.2}
\begin{tabular}{ c c c c c c c}
\hline
\multirow{2}{*}{Case} & \multirow{2}{*}{Excitation}  & \multicolumn{3}{c}{Amplitude} & \multicolumn{2} {c} {Phase}  \\
 &  & Pitch $\alpha_{tot}$ [deg] & Heave $\dot{h}_0/U$ [deg] & Gust $w_0/U$ [deg] &$\Delta \phi_{w,\alpha}$ [deg] & $\Delta \phi_{w,\dot{h}}$ [deg]\\
\hline
$\mathrm{I}$ - $\alpha_{e}^{\mathrm{I}}$ & Pitch & 1.5 & 0.0 & 0.0 & 0&0\\
$\mathrm{II}$ - $\alpha_{e}^{\mathrm{II}}$ & Pitch & 4.0 & 0.0 & 0.0 & 0&0\\
$\mathrm{III}$ - $\alpha_{e}^{\mathrm{III}}$  & Heave & 0.0 & 1.5 & 0.0 & 0&0\\
$\mathrm{IV}$ - $\alpha_{e}^{\mathrm{IV}}$  & Heave & 0.0 & 4.0 & 0.0 & 0&0\\
$\mathrm{V}$ - $\alpha_{e}^{\mathrm{V}}$  & Gust & 0.0 & 0.0 & 1.5 & 0&0\\
$\mathrm{VI}$ - $\alpha_{e}^{\mathrm{VI}}$  & Gust & 0.0 & 0.0 & 4.0 & 0&0\\
$\mathrm{VII}$ - $\alpha_{e}^{\mathrm{VII}}$  & Pitch + Gust & 1.5 & 0.0 & 1.5 & 0&0\\
$\mathrm{VIII}$ - $\alpha_{e}^{\mathrm{VIII}}$  & Pitch + Gust & 4.0 & 0.0 & 4.0 & 0&0\\
$\mathrm{IX}$ - $\alpha_{e}^{\mathrm{IX}}$  & Pitch + Gust & 1.5 & 0.0 & 1.5 & 90&0\\
$\mathrm{X}$ - $\alpha_{e}^{\mathrm{X}}$  & Pitch + Gust & 4.0 & 0.0 & 4.0 & 90&0\\
$\mathrm{XI}$ - $\alpha_{e}^{\mathrm{XI}}$  & Heave + Gust & 0.0 & 1.5 & 1.5 & 0&0\\
$\mathrm{XII}$ - $\alpha_{e}^{\mathrm{XII}}$  & Heave + Gust & 0.0 & 4.0 & 4.0 & 0&0\\
$\mathrm{XIII}$ - $\alpha_{e}^{\mathrm{XIII}}$  & Heave + Gust & 0.0 & 1.5 & 1.5 & 0&90\\
$\mathrm{XIV}$ - $\alpha_{e}^{\mathrm{XIV}}$  & Heave + Gust & 0.0 & 4.0 & 4.0 & 0&90\\
\hline
\end{tabular}
\caption{{ Study cases based on the effective angle of attack, constituted from $\alpha_{tot}$, $\dot{h}/U$ and $w/U$. }}
\label{T1}
\end{table}

\subsection{Computational Fluid Dynamics Model}
\label{VP}
The CFD model is based on the VPM to simulate the fluid-structure interaction in the CFD domain that is depicted in Fig.~\ref{Fig2}. Herein, the governing equations and the numerical implementation of the VPM are only briefly revisited, while more information can be found in \citep{Cottet2000,Ge2008,Morgenthal2002,Larsen1997}

The fluid is governed by the fundamental two-dimensional (2D) Navier–Stokes (NS) equations for incompressible viscous fluid. In the vorticity transport form, these equations are described as: 
\begin{equation}
\frac{\partial \omega_u}{\partial t}+(\boldsymbol{u} \cdot \nabla) \boldsymbol{\omega}_u=\nu \nabla^{2} \boldsymbol{\omega}_u,
\end{equation}
where $\boldsymbol{u}(\boldsymbol{x},t)$ is the velocity vector at postion $\textbf{\textit{x}}$,  $\nu$ is the kinematic viscosity and $\boldsymbol{\omega}_u$ is vorticity defined as $\boldsymbol{\omega}_u=\nabla \times \boldsymbol{u}$.
Using the inverted kinematic relation, the velocity field can be obtained from the vorticity field as:
\begin{equation}
\label{posn}
\nabla^{2} \boldsymbol{u}=-\nabla \times \boldsymbol{\omega}_u,
\end{equation}
The preceding equation represents a Poisson equation and can be solved using Green's function, yielding the Biot-Savart relation between velocity and vorticity:
\begin{equation}
\label{BS}
\boldsymbol{u}(\boldsymbol{x})=\boldsymbol{U}-\frac{1}{2 \pi} \int_{\mathcal{D}} \frac{ \left(\boldsymbol{x}-\boldsymbol{z}\right)\times\boldsymbol{\omega}_u\left(\boldsymbol{z}\right)}{\left|\boldsymbol{x}-\boldsymbol{z}\right|^{2}} \mathrm{~d} \boldsymbol{z},
\end{equation}
where $\boldsymbol{U}$ is the free-stream velocity vector and $\mathcal{D}$ is the entire domain comprising the fluid and immersed body. Equation~\eqref{BS} represents the basic equation for numerical discretisation in the VPM.\par
In the VPM, the vorticity ﬁeld is represented by a ﬁnite number of particles $N_p$ characterized by their locations $\boldsymbol{x}_{p}$ and strength $\boldsymbol{\Gamma_{p}}$. The strength of the vortices in 2D is a scalar and is obtained by integrating the vorticity over a small patch of fluid as $\Gamma_{p}=\int_{\mathcal{D}_{p}} 
{\omega}_u(\boldsymbol{x}) d \mathcal{D}_{p}$. Substituting for the vorticity in Eq.~\eqref{BS}, the Biot-Savart relation can be obtained as
\begin{equation}
\label{BS2}
\boldsymbol{u}\left(\boldsymbol{x}_{p}\right)=\boldsymbol{U}-\frac{1}{2 \pi} \sum_{p=1}^{N_{p}} \frac{\left(\boldsymbol{x}-\boldsymbol{x}_{p}\right) \times \boldsymbol{\Gamma}_{p}}{\left|\boldsymbol{x}-\boldsymbol{x}_{p}\right|^{2}}=\boldsymbol{U}-\sum_{p=1}^{N_{p}} \boldsymbol{K} \Gamma_{p},
\end{equation}
where $\boldsymbol{\Gamma}_{p}=\Gamma_{p} \boldsymbol{e_z}$ for unit vector $\boldsymbol{e}_{z}$ and $\boldsymbol{K}=\boldsymbol{K}\left(\boldsymbol{x}-\boldsymbol{x}_{p}\right)$ is a velocity kernel, that is subsituted with a mollified velocity kernel $\boldsymbol{K}_\epsilon$ to account for the numerical instabilities if $|\boldsymbol{x}-\boldsymbol{x}_p|\rightarrow0$.\par
The current VPM solver is the one of \cite{Morgenthal2007}, which employs the Particle–Particle-Particle–Mesh (P$^3$M) algorithm for the solution of Eq.~\eqref{BS2} to increase computational efficiency. Furthermore, the two-step viscous splitting technique is employed to solve the vorticity transport equation into a convection and a diffusion step. In the convection step, the particles are first convected using the standard Runge-Kutta time-marching techniques. In the diffusion step, the particles are randomly perturbed using the stochastic Random Walk Method. The no-slip and no-penetration boundary conditions are implicitly enforced by the production of vorticity on the surface of the structure (vortex sheet). The section forces are then calculated by integrating the surface pressures, as $\nabla p=\mu \nabla \times \boldsymbol{\omega}_u$ where $\mu$ is the laminar viscosity.\par
\begin{figure}[h!]
\includegraphics[trim=0cm 0cm 0cm 0.1cm,clip,width=1\columnwidth]{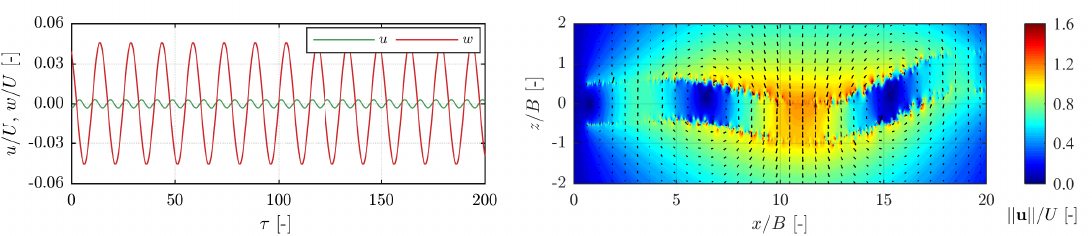} 
\caption{Sinusoidal vertical gust generation for the CFD model: Velocity fluctuations at tracking point $x/B = $ 6 and $z/B$=0 (left), and instantaneous velocity field in the computational domain (right).  (For interpretation of the references to color in this figure legend, the reader is referred to the web version of this article.) }
\label{Velocity}
\end{figure}

The method presented by \cite{Kavrakov2019} is employed to simulate sinusoidal vertical gusts in the CFD domain. The method rests on the concept of Active Turbulence Generator used in wind tunnel experiments~\cite{Diana2013}, and it mimics the wake flow of two sinusoidally pitching fictitious airfoils $F_{A}$ and $F_{B}$ that are positioned upstream of the section. The phase between the airfoil motion $\alpha_ {F_A}$ and $\alpha_{F_B}$ dictate the direction of the generated gust. In-phase oscillation generates vertical gust while out-of-phase generates horizontal gust. The wakes of the airfoils are replicated in the CFD domain by seeding two particles at the $j^{th}$ time step at location 1 and 2 that are separated by distance $R$. The particles carry concentrated circulation $\Gamma_{F_A,j}$ $\Gamma_{F_B,j}$, which induces a sinusoidal single frequency vertical gust with amplitude $w_{ 0}$  along the center axis $x$.  A closed-form solution is given in~\citep{Kavrakov2019} relates the particle circulation with the target gust amplitude. A closed-form solution is given in~\citep{Kavrakov2019} relates the particle circulation with the target gust amplitude. As an example, Fig.~\ref{Velocity} shows the velocity fluctuations for non-dimensional time $\tau=tU/B$ and the corresponding instantaneous ﬁelds of ﬂuctuating velocity magnitude $||$\textbf{u}$||$, for in-phase oscillation of the fictitious airfoils, yielding a vertical sinusoidal gust.

\subsection{Linear Unsteady Model}
\label{LU_FP}

The aerodynamic forces for the system shown in Fig.~\ref{Fig2} can be described using the LU model. Originally developed by Davenport and Scanlan \citep{Scanlan1971a,Davenport1962}, the LU model splits the contribution of the gust- and motion-induced forces acting on a body as:
\begin{equation}
L_{\mathrm{LU}}=L_{se}+L_b, 
\; \;  \; \;  
M_{\mathrm{LU}}=M_{se}+M_b, 
\end{equation}
where $L_{se}$ and $M_{se}$ are the motion-induced (self-excited) lift and moment forces, respectively, while $L_b$ and $M_b$ are the gust-induced (buffeting) lift and moment forces, respectively. The self-excited forces in extended Scanlan's format are given as:
\begin{equation}
 \label{LUse}
\begin{split}
%D_{se}=\frac{1}{2} \rho U^{2} B\left(K P_{1}^{\ast} \frac{\dot{p}}{U}+K P_{2}^{\ast} %%\frac{B \dot{\alpha}}{U}+K^{2} P_{3}^{\ast} \alpha +K^{2} P_{4}^{\ast} \frac{p}{B}+ K P_{5}^{\ast} \frac{\dot{h}}{U} + K^{2} P_{6}^{\ast} \frac{h}{B}\right) 
%\\
L_{se}=\frac{1}{2} \rho U^{2} B\left(K H_{1}^{\ast} \frac{\dot{h}}{U}+K H_{2}^{\ast} \frac{B \dot{\alpha}}{U}+K^{2} H_{3}^{\ast} \alpha+K^{2} H_{4}^{\ast} \frac{h}{B} \right),
\\
M_{se}=\frac{1}{2} \rho U^{2} B^{2}\left(K A_{1}^{\ast} \frac{\dot{h}}{U}+K A_{2}^{\ast} \frac{B \dot{\alpha}}{U}+K^{2} A_{3}^{\ast} \alpha+K^{2} A_{4}^{\ast} \frac{h}{B} \right),
\end{split}
\end{equation}
where $\rho$ is the air density, $H_{j}^{\ast}(K)$ and $A_{j}^{\ast}(K)$ for $j \in \{1,...,4 \}$ are the aerodynamic derivatives that are fucntion of the reduced frequency $K=B \omega /U$, for $\omega$ being circular frequency of the motion. The reduced frequency can be also related to the reduced velocity $V_r=U/(fB)=2\pi/K$. The unsteady buffeting force is expressed based on Davenport’s general framework,
\begin{equation}
 \label{LUb}
\begin{split}
%D_{b}=\frac{1}{2} \rho U^{2} B\left[2 C_{D} \chi_{D u} \frac{u}{U}+\left(C_{D}^{\prime}-C_{L}\right) \chi_{D w} \frac{w}{U}\right]
%\\
L_{b}=-\frac{1}{2} \rho U^{2} B\left[2 C_{L} \chi_{L u} \frac{u}{U}+\left(C_{L}^{\prime}+C_{D}\right) \chi_{L w} \frac{w}{U}\right],
\\
M_{b}=\frac{1}{2} \rho U^{2} B^{2}\left(2 C_{M} \chi_{M u} \frac{u}{U}+C_{M}^{\prime} \chi_{M w} \frac{w}{U}\right),
\end{split}
\end{equation}
where $\chi_{j u} (K)$ and $\chi_{j w} (K) $  are the aerodynamic admittance functions,  $ C_{j}(\alpha_{s})$ and $ C_{j}^\prime=C_{j}^\prime(\alpha_{s})|_{\alpha_{s}}$ are the static aerodynamic coefficients and their derivatives with respect to the static angle of attack $\alpha_{s}$ for $j \in \{ L, M, D\}$. The aerodynamic admittance functions are complex functions in the form $ \chi = F + iG$, where $F$ and $G$ represent the magnitude and phase ratios, respectively, between the aerodynamic forces and the free-stream gust.\par
The aerodynamic derivatives are determined using CFD and the forced vibration technique as discussed in \citep{Larsen1998}. The complex form of the aerodynamic admittance functions is computed as a transfer function between the sinusoidal free-stream gust and the corresponding buffeting force acting on the bluff body, as presented in \cite{Kavrakov2019}. Using this method, the gust is first tracked down in the CFD domain in simulation without a cross-section (i.e. fluid only). Then in another simulation, the cross-section is placed in the CFD domain (i.e. fluid-structure interaction) and is subjected to a sinusoidal vertical gust so that the gust-induced forces are measured. Having the gust time-history and the corresponding aerodynamic force, the complex form of aerodynamic admittance can be determined from the phase and amplitude modulations based on Eq.~\eqref{LUb}. This way of computing the complex form of admittance is only possible because the horizontal gust component $u$ is negligible.

\subsection{Comparison Metrics for Force Time-histories}
 \label{metrics}
Advanced comparison metrics for time-histories are employed~\citep{Kavrakov2020} to compare the force time-histories of the CFD and LU models. Using these metrics, it is possible to delineate particular features in the combined forces of the CFD model due to the nonlinear interaction between the gust and motion.  Such nonlinear features may include the high-frequency content that alludes to vortex shedding and higher-order frequencies in the forces that are due to the unsteady flow separation and reattachment (shear layer instability) and wake instability~\citep{Wu2013b}. A comparison metric $\mathcal{M}^{a,b}$ is constructed to compare the signals $a=a(t)$ and $b=b(t)$, where $a$ is the reference. For the present study, the CFD model is always taken as the reference. Each metric can take a value from 0 and 1, where 1 is a perfect match for a particular signal feature. The employed set of metrics includes a phase $\mathcal{ M}_{\varphi}$,  peak $\mathcal{ M}_{p}$, root-mean-square (RMS) $\mathcal{ M}_{rms}$, wrapped magnitude $\mathcal{ M}_{m} $, wavelet $\mathcal{ M}_{w} $, frequency-normalized wavelet $\mathcal{M}_{wf}$, and a bispectrum $\mathcal{ M}_{b}$ metric. 
The phase metric $\mathcal{ M}_{\varphi}$  quantifies the average phase shift between the two signals. This measures the quality of the unsteady aerodynamic coefficients to represent the time-lag between the input signals and the induced forces. The peak $\mathcal{ M}_{p}$ and RMS $\mathcal{ M}_{rms}$ metrics evaluate the average magnitude discrepancies, while the magnitude $\mathcal{ M}_{m} $ metric evaluates the local magnitude differences. When calculating $\mathcal{ M}_{m} $,  first the two signals are shifted, but not scaled, to alleviate the effects of the local phase shifts and high-frequency contents by using the dynamic time warping (DTW) algorithm. The wavelet-based $\mathcal{ M}_{w}$ is applied to capture the spectral distortion in the time-frequency plane. The distortion can be ascribed to amplitude and/or frequency modulation.  To distinguish that, the frequency marginalized $\mathcal{ M}_{wf}$ metrics, which individually represent the frequency modulation, is employed. Finally, a wavelet-based bispectrum $\mathcal{ M}_{b}$ is introduced to capture quadratic nonlinearities. In addition to the comparison metrics, the peak spectral amplitude ratio of the forces used to characterize the accuracy of the LU model for a selected $V_r$ range.

\section{Linear Behaviour: Flat Plate}
\label{Sec3_Verif}
\begin{figure}[h!]
\centering
\includegraphics[trim=0cm 0.1cm 0cm 0.1cm]{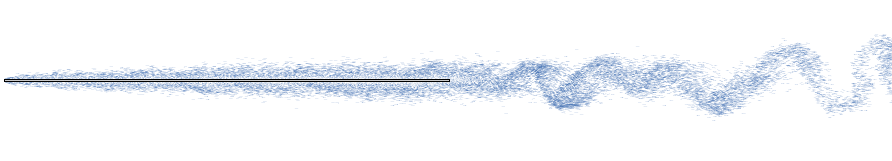} 
\caption{Particle map of a flat plate with an aspect ratio of $B/D=200$ at $\mathrm{Re}=1\times10^4$.}
\label{FP_Section}
\end{figure}

The CFD procedure is first verified for the analytical solution for the LU model of a flat plate by comparing the aerodynamic derivatives and admittance with their analytical counterpart by Sears~\citep{Davenport1962} and Theodorsen~\citep{Scanlan1971a} at Reynolds number $\mathrm{Re}=1\times10^4$. The CFD domain is set as  $-11B < L_{x} < 11B$ and $-5.5B < L_{z} < 5.5B$  (see Fig.~\ref{Fig2}) and airfoil distance $R/B$ = 1.5 for gust simulation. The section is discretized on 402 panels with an average length amounting to $ \Delta l /B$= $2.5$ x $10^{-3}$ and a reduced time-step of $\Delta t U/B$ = $5\times10^{-3}$. Similar CFD parameters were selected in \citep{Kavrakov2019a}, where the boundary layer was verified as well. Figure~\ref{FP_Section} depicts an instantaneous particle map. 
%$\tau=\Delta t U/B$ = $5$ x $10^{-3}$

\begin{figure}[h!]
\centering
\includegraphics[trim=0cm 0cm 0cm 0cm,clip,width=1\columnwidth]{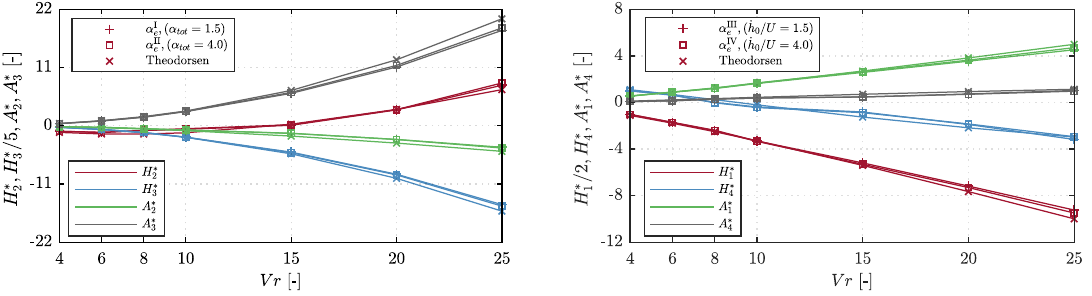} 
%\vspace*{-0.5cm}
\caption{Flat plate aerodynamic derivatives related to the pitch (left) and heave DOF (right). Angle values are in [deg].  (For interpretation of the references to color in this figure legend, the reader is referred to the web version of this article.) }
\label{FD_FP}
\end{figure}

\begin{figure}[h!]
\centering
\includegraphics[trim=0cm 0cm 0cm 0cm,clip,width=1\columnwidth]{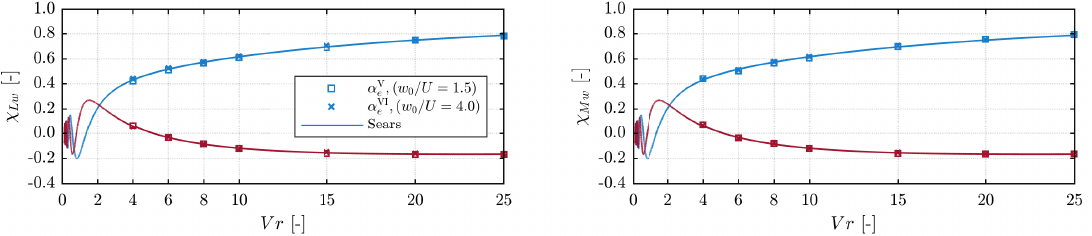} 
%\vspace*{1cm}
\caption{Flat plate aerodynamic admittance of the lift (left) and moment (right) due to deterministic vertical gust:  Real (blue) and imaginary parts (red). Angle values are in [deg].  (For interpretation of the references to color in this figure legend, the reader is referred to the web version of this article.) }
\label{ADD_FP}
\end{figure}

%\FloatBarrier
Figure \ref{FD_FP} and \ref{ADD_FP} respectively depict the simulated aerodynamic derivative and aerodynamic admittance plots along with the analytical solution, where $Vr$ is the reduced velocity. Reduced velocities $V_r$ ranging from 4 to 25 considered to represent both the steady and unsteady range for two amplitudes. Good agreement is achieved between the analytical and CFD values for both amplitude cases. Assessing the aerodynamic coefficients of the two amplitude cases shows the individual self-excited or buffeting forces are linear in the selected amplitude range.

%\vspace*{-0.2cm}

\begin{figure}[h!]
\centering
\includegraphics[trim=0cm 0cm 0cm 0cm,clip,width=1\columnwidth]{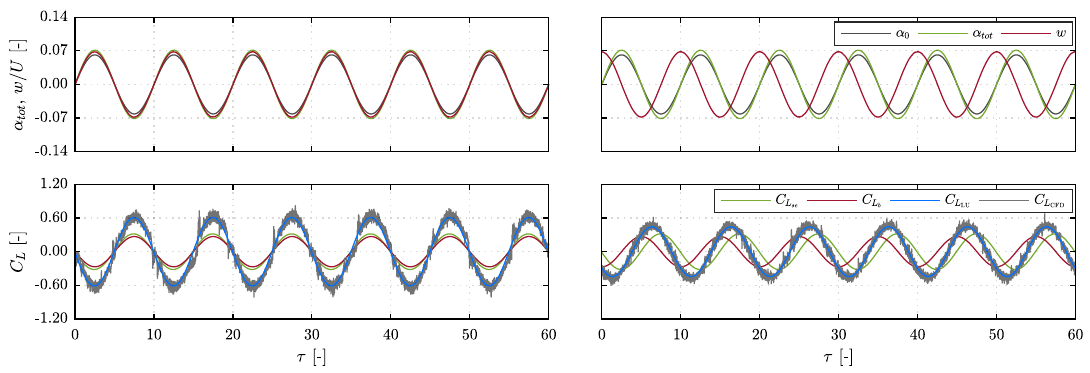} 
\vspace*{-0.6cm}
\caption{Flat plate: Input pitch motion and gust (top), with the corresponding fluctuating lift $C_L$ coefficient (bottom) for the CFD and LU model for combined forcing (gust and motion) at $V_r$ = 10. The separate contribution of the motion- ($C_{L_{se}}$) and gust-induced forces ($C_{L_{b}}$) are those of the LU model ($C_{L_{\mathrm{LU}}}=C_{L_{se}}+C_{L_{b}}$). Case $\alpha_{e}^{\mathrm{VIII}}$ ($\alpha_{tot}$=4.0 deg with $\alpha_0$=3.5 deg, $w_0/U$= 4.0 deg, $\Delta\phi_{w,\alpha}$=0 deg) left and case $\alpha_{e}^{\mathrm{X}}$ ($\alpha_{tot}$=4.0 deg with $\alpha_0$=3.5 deg, $w_0/U$= 4.0 deg,  $\Delta\phi_{w,\alpha}$=90 deg) right.  (For interpretation of the references to color in this figure legend, the reader is referred to the web version of this article.)  }
\label{Force_FP_Vr10_Pitch}
\end{figure}

\begin{figure}[h!]
\centering
\includegraphics[trim=0cm 0cm 0cm 0cm,clip,width=1\columnwidth]{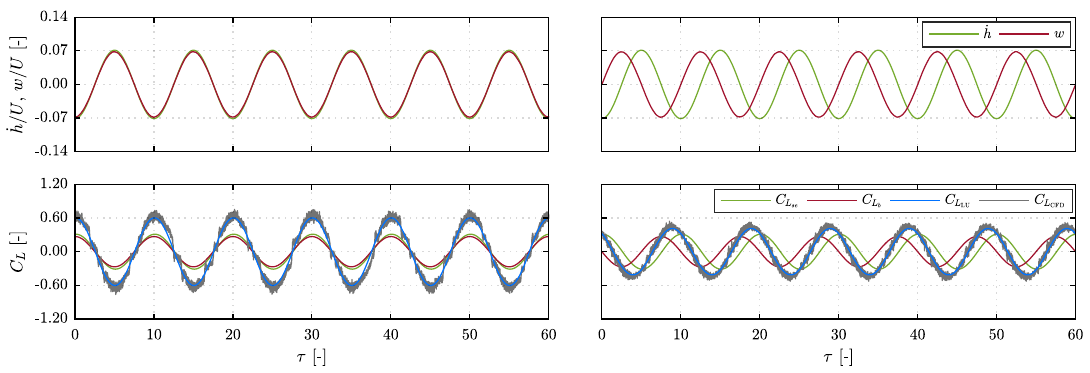} 
\vspace*{-0.6cm}
\caption{Flat plate: Input heave motion and gust (top), with the corresponding fluctuating lift $C_L$ coefficient (bottom) for the CFD and LU model for combined forcing (gust and motion) at $V_r$ = 10. The separate contribution of the motion- ($C_{L_{se}}$) and gust-induced forces ($C_{L_{b}}$) are those of the LU model ($C_{L_{\mathrm{LU}}}=C_{L_{se}}+C_{L_{b}}$). Case $\alpha_{e}^{\mathrm{XII}}$ ($\dot{h}/B$=4.0 deg, $w_0/U$= 4.0 deg, $\Delta\phi_{w,\dot{h}}$=0 deg) left and case $\alpha_{e}^{\mathrm{XIV}}$ ($\dot{h}/B$=4.0 deg, $w_0/U$= 4.0 deg,  $\Delta\phi_{w,\dot{h}}$=90 deg) right.  (For interpretation of the references to color in this figure legend, the reader is referred to the web version of this article.)  }
\label{Force_FP_Vr10_Heave}
\end{figure}

\begin{figure}[h!]
\centering
\includegraphics[trim=0cm 0.5cm 0cm 0cm,clip,width=1\columnwidth]{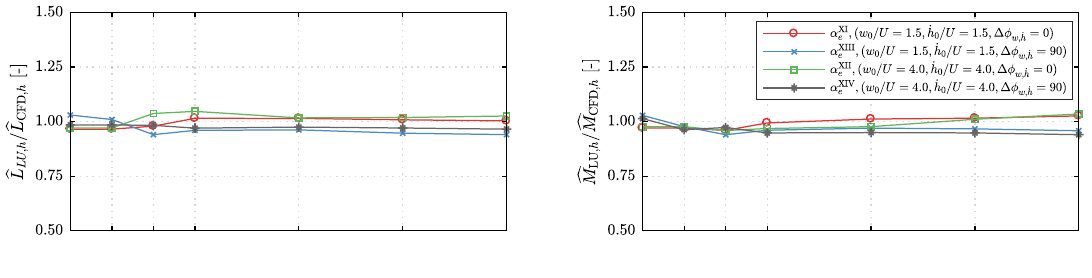} 
\includegraphics[trim=0cm 0cm 0cm 0cm,clip,width=1\columnwidth]{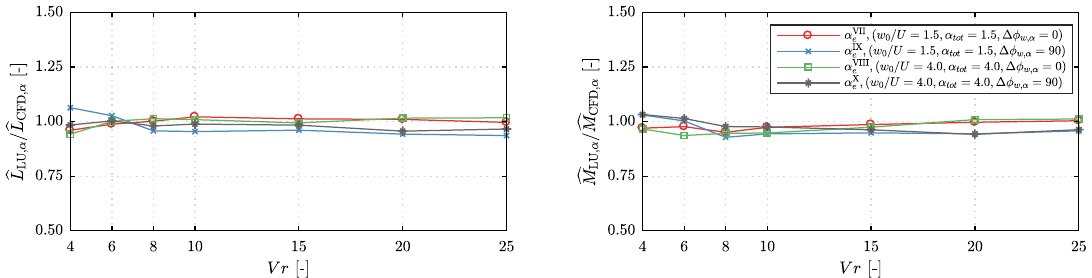}
\vspace*{-0.5cm}
\caption{Flat plate: Peak aerodynamic force spectral amplitude ratio for the LU and CFD models. The aerodynamic forces due forced motion inputs in the heave (top) and pitch (bottom) DOF. Angle values are in [deg].  }
\label{fp_SpRMS_Hv}
\end{figure}

A sinusoidal gust is then applied on the oscillating flat plate section for the cases XI, XIII, XII, and XIV (see Tab.~\ref{T1} and Fig.~\ref{Fig2}).  Figure \ref{Force_FP_Vr10_Pitch} and \ref{Force_FP_Vr10_Heave, respectively, depict the simulated and LU model lift force coefficient for pitch and heave oscillation with vertical gust at $V_r$=10, with and without phase shift (cases VII, X, XII, and XIV). For both DOFs and phase inputs, the CFD and LU model have  very good agreement.} The peak spectral amplitude ratios in the Fast Fourier Transforms (FFTs) shown in Fig.~\ref{fp_SpRMS_Hv} complement this observation, where $\widehat{L}_{jk}$ and $\widehat{M}_{jk}$ represent the peak lift and moment spectral amplitude ratio for $j \in$  \{$\mathrm{LU},\mathrm {CFD}$\} and $k \in$ \{$h,\alpha$\}, respectively. This shows that the linear hypothesis is valid for a flat plate in the selected amplitude range and is accurately predicted by the CFD model.

\vspace*{-0.5cm}

%\FloatBarrier

\section{Nonlinear Behaviour: Bridge Decks}
\label{Sec4_Appl}
\begin{figure}[h!]
\centering
\begin{minipage}{.5\textwidth}
  \centering
  \includegraphics[width=0.8\linewidth]{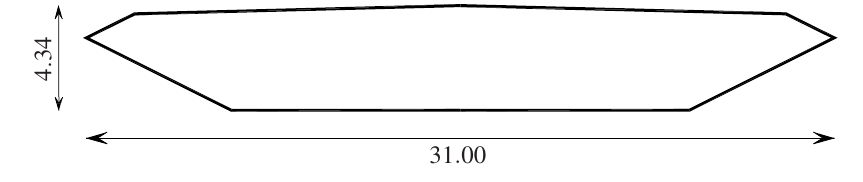}
\end{minipage}%
\begin{minipage}{.5\textwidth}
  \centering
    \includegraphics[width=0.8\linewidth]{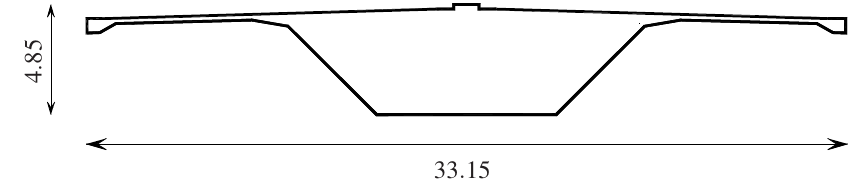}
\end{minipage}
\caption{Streamlined deck of the Great Belt Bridge (left) and bluff generic box girder deck (right). Dimensions in [m]. }
 \label{Xsection_MG_GB}
\end{figure}

Having the CFD procedure verified, this section studies the combined behaviour of the gust- and motion-induced forces for two generic bridge decks shown in Fig.~\ref{Xsection_MG_GB}. The first one is the streamlined deck of the Great Belt Bridge, while the second one is a bluff generic box girder. In the discussion, these two decks are termed as streamlined and bluff deck, respectively. 
Although the decks' aspect ratios are roughly the same, they experience different flow dynamics in the shear layer. For the selected amplitude range, the streamlined deck exhibits a slight flow separation. On the contrary, the bluff deck displays prominent flow separation from the bottom sharp-edged corners, which is more sensitive to the incident angle of attack due to its asymmetric geometry. The CFD studies are conducted at a Reynolds Number of $\mathrm{Re}=1\times10^5$, with reduced time-step  $\Delta t U/B$ = $1.65$ x $10^{-2}$ . The decks are discretized on 250 panels with average panel length of $\Delta l /B$= $8.2$ x $10^{-3}$. The CFD parameters for the Great Belt deck are the ones of \cite{Kavrakov2018b}, wherein the CFD results were compared with other studies yielding good correspondence.\par 

In what follows, the individual components in the aerodynamic forces due to gust and motion are studied first, and the linear hypothesis is investigated through the aerodynamic derivatives and admittances. Then, the nonlinear interaction between these force components is studied through their nonlinear dependence on the combined angle of attack (gust and motion) and the influence of the free-stream gusts on the shear layer of a moving deck.

\begin{figure}[!t]
	\centering
	\includegraphics[trim=0cm 0.5cm 0cm 0cm,clip,width=1\columnwidth]{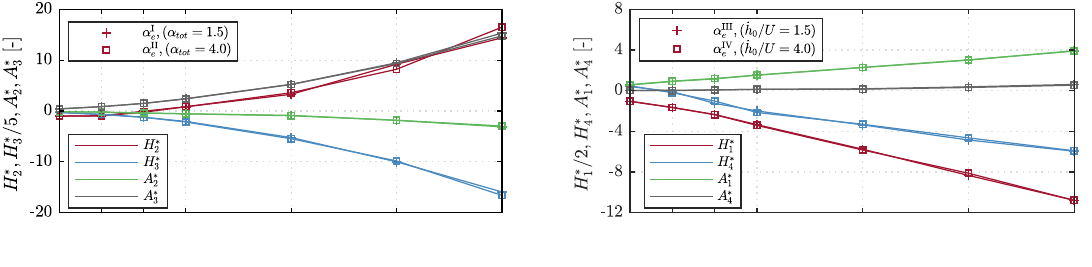} 
	%\vspace*{-0.5cm}
	\includegraphics[trim=0cm 0.1cm 0cm 0cm,clip,width=1\columnwidth]{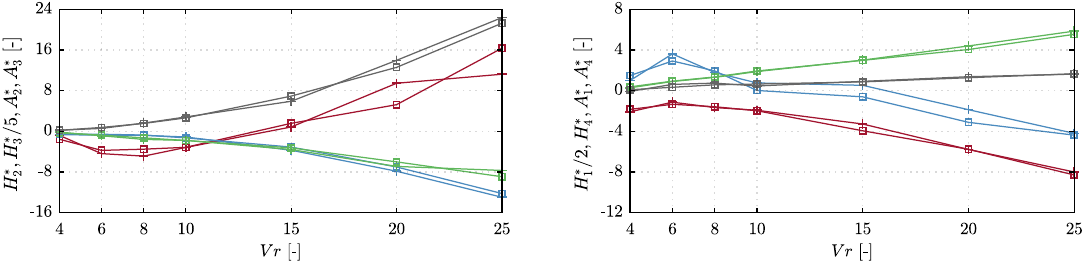} 
	\vspace*{-0.6cm}
	\caption{Streamlined (Top) and bluff deck's (bottom) aerodynamic derivatives.  Left: related to the pitch DOF, right: related to the heave DOF. Angle values are in [deg].  (For interpretation of the references to color in this figure legend, the reader is referred to the web version of this article.) }
	\label{FD_GB_MG}
\end{figure}
\begin{figure}[!t]
	\centering
	\includegraphics[trim=0cm 0.5cm 0cm 0.00cm,clip,width=1\columnwidth]{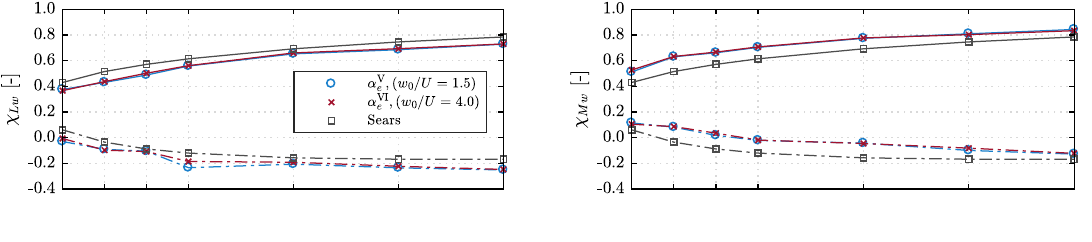} 
	%\vspace*{-0.5cm}
	\includegraphics[trim=0cm 0.05cm 0cm 0cm,clip,width=1\columnwidth]{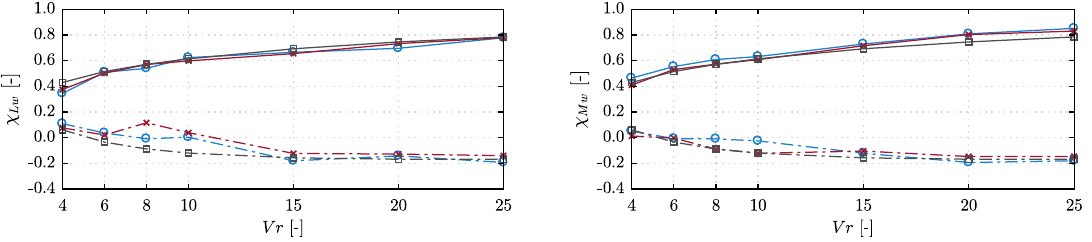} 
	\vspace*{-0.6cm}
	\caption{Streamlined (Top) and bluff deck's (bottom) aerodynamic admittance for deterministic vertical gust. Real (solid line) and imaginary  (broken line) parts. Angle values are in [deg].  }  
	\label{GB_MG_ADD}
\end{figure}
\FloatBarrier
\subsection{Aerodynamic forces due to separate effects of gust and motion}
Figure \ref{FD_GB_MG} and \ref{GB_MG_ADD} show the simulated aerodynamic coefficients of both decks for two amplitude cases. Good agreement can be observed for the aerodynamic derivatives for the two amplitude cases. Minor differences can be observed for the bluff deck's $H_{2}^{\ast}$ and $H_{4}^{\ast}$ derivatives, which may be attributed to the interior noise in the aerodynamic coefficients.  A previous study \citep{Kavrakov2018b} has shown that the flutter derivatives for the streamlined deck of the current CFD method have a good agreement with experimental results. The bluﬀ section’s ﬂutter derivatives also show a similar trend with experimental and CFD results from generic box-girders with similar properties \citep{Mannini2016,Xu2017}.  The aerodynamic admittances of the bridge decks are depicted in Fig.~\ref{GB_MG_ADD}. These are determined following the procedure presented in Sec.~\ref{LU_FP}. Generally, similar results are obtained for both amplitude cases for the streamlined deck, with minor differences in the phase value for the bluff deck. Using deterministic sinusoidal gusts resembles an actively generated turbulence in experimental testing. With this procedure (experimental or numerical), the 2D aerodynamic admittance is obtained (see \cite{Kavrakov2018b} for discussion); hence, the admittances are comparable to Sears' admittance function. The body-generated turbulence dominates the response at low reduced velocities; thus, a cut-off of $V_r=2$ is considered. The good agreement between the unsteady aerodynamic coefficients (derivatives and admittances) for the two amplitude cases yields the reasoning that the individual motion- and gust-induced force components are linear in the main harmonic for the selected amplitude and $V_r$ range of both deck sections.\par

\begin{figure}[h!]
\centering
\includegraphics[trim=0cm 0cm 0cm 0cm,clip,width=1\columnwidth]{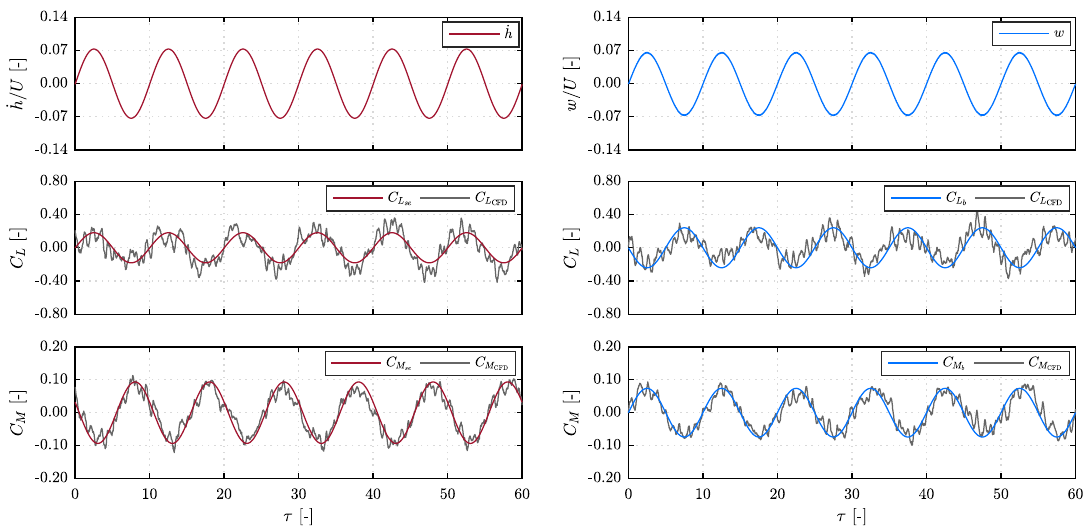} 
%\vspace*{-0.7cm}
\caption{Bluff deck: Input motion and gust (top), with the corresponding fluctuating lift $C_L$ (center) and moment $C_M$ (bottom) coefficients for the CFD and LU model for separate forcing at $V_r$ = 10. The separate contribution of the motion- ($C_{L_{se}}$ and $C_{M_{se}}$) and gust-induced forces ($C_{L_{b}}$ and $C_{M_{b}}$) are those of the LU model. Case $\alpha_{e}^{\mathrm{IV}}$ ($\dot{h}_0/U$=4.0 deg) left and case $\alpha_{e}^{\mathrm{VI}}$ ($w_0/U$= 4.0 deg) right.  (For interpretation of the references to color in this figure legend, the reader is referred to the web version of this article.) }
\label{Force_MG_Vr10_HG}
\end{figure}
%\vspace*{-0.6cm}
\begin{figure}[h!]
\centering
\includegraphics[trim=0cm 0.0cm 0cm 0.0cm,clip,width=1\columnwidth]{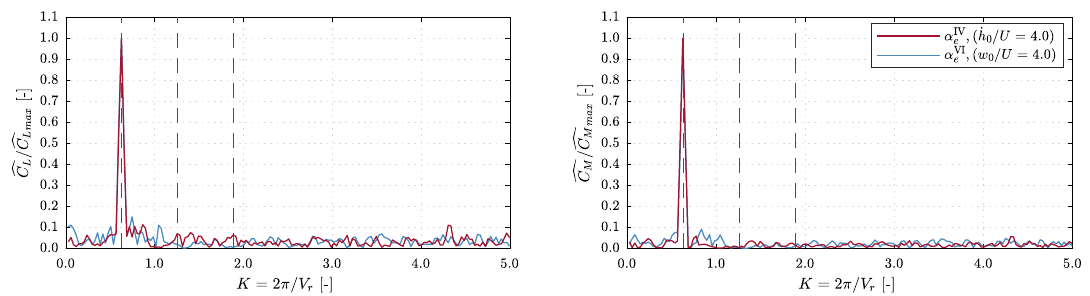} 
\vspace*{-0.6cm}
\caption{Bluff deck: Normalized FFT of the lift $C_L$ (left) and moment $C_M$ (right)  coefficients for the CFD  model of the separate forcing cases at $V_r$=10. The broken vertical lines denote the main, the second, and the third harmonics $K$ values.  Angle values are in [deg].  (For interpretation of the references to color in this figure legend, the reader is referred to the web version of this article.) }
\label{spectralPlot_vr_10_HG}
\end{figure}
Figures~\ref{Force_MG_Vr10_HG} and~\ref{spectralPlot_vr_10_HG} show the bluff decks time-history and normalized FFT of the CFD forces due to separate input motion and gust excitation cases ($\mathrm{IV}:\dot{h}_0/U=4$ deg, $\mathrm{VI}:w_0/U=4$ deg) at $V_r=10$. The aerodynamic forces for the LU model are practically the linear curve fit of the CFD force, based on the aerodynamic derivatives and admittances. No higher-order  harmonics appear for both cases, which are a typical indication of aerodynamic nonlinearity. Table~\ref{T2} lists the relative contribution of the second and third-order harmonics in the aerodynamic forces with respect to the main harmonic for the high-amplitude case. A threshold of 10\% is roughly to distinguish between potential nonlinearity and spectral noise. The small amplitudes cases (I, III, V, VII, IX, XI, XIII) are not included due to the high vortex shedding and interior noise resulting in noise in the FFT. Generally, the results show that the higher-order harmonics in the aerodynamic forces of the separate cases are not significant, with a slightly higher contribution for the two motion cases ($\mathrm{II}:\alpha_{tot}=4$ deg at $V_r$=4 and $\mathrm{IV}:\dot{h}_0/U=4$ at $V_r$=25, Tab.~\ref{T1}). However, this increase in the higher harmonics is near the threshold and may be attributed to the higher signal-to-noise ratio in the time-histories for $V_r=4$ (not shown). Thus, the aerodynamic nonlinearity manifested through the higher-order harmonics is insignificant for the separate excitation due to gust or motion. 
\FloatBarrier

\subsection{Aerodynamic forces due to combined effects of gust and motion}
\label{Result_1}

\begin{figure}[h!]
\centering
\includegraphics[trim=0cm 0.05cm 0cm 0cm,clip,width=1\columnwidth]{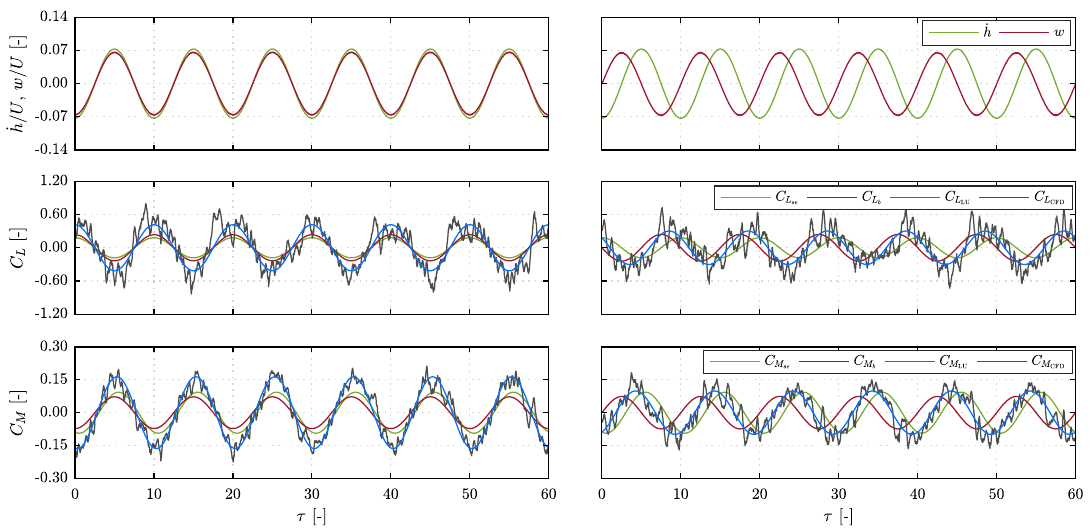} 
%\vspace*{-0.7cm}
\caption{Bluff deck: Input motion and gust (top), with the corresponding fluctuating lift $C_L$ (center) and moment $C_M$ (bottom) coefficients for the CFD and LU model for combined forcing (gust and motion) at $V_r$ = 10.  The separate contribution of the motion- ($C_{L_{se}}$ and $C_{M_{se}}$) and gust- ($C_{L_{b}}$ and $C_{M_{b}}$) induced forces  are those of the LU model ($C_{L_{\mathrm{LU}}}=C_{L_{se}}+C_{L_{b}}$ and $C_{M_{\mathrm{LU}}}=C_{M_{se}}+C_{M_{b}}$). Case $\alpha_{e}^{\mathrm{XII}}$ ($\dot{h}_0/U$=4.0 deg, $w_0/U$= 4.0 deg, $\Delta\phi_{w,\dot{h}}$=0 deg) left and case $\alpha_{e}^{\mathrm{XIV}}$ ($\dot{h}_0$=4.0 deg $w_0/U$= 4.0 deg,  $\Delta\phi_{w,\dot{h}}$=90 deg) right. (For interpretation of the references to color in this figure legend, the reader is referred to the web version of this article.) }
\label{Force_MG_Vr10_Heave}
\end{figure}

Next, both decks are subjected to a combined action of motion and gust. The aerodynamic admittance and flutter derivatives determined in the previous are used as an input for the LU model to establish the baseline for the linear superposition principle. Figure~\ref{Force_MG_Vr10_Heave} depicts sample time-histories for the bluff deck of the input motion and gusts (top) at $V_r$=10 and corresponding aerodynamic forces for two cases (XII and XIV, $h_0/U$=$w_0/U$=4 deg) with the same amplitude of the separate gust and heave contributions to the effective angle of attack, however, with a phase difference.

%\vspace*{-0.6cm}
\begin{figure}[h!]
\centering
\includegraphics[trim=0cm 0.0cm 0cm 0.1cm,clip,width=1\columnwidth]{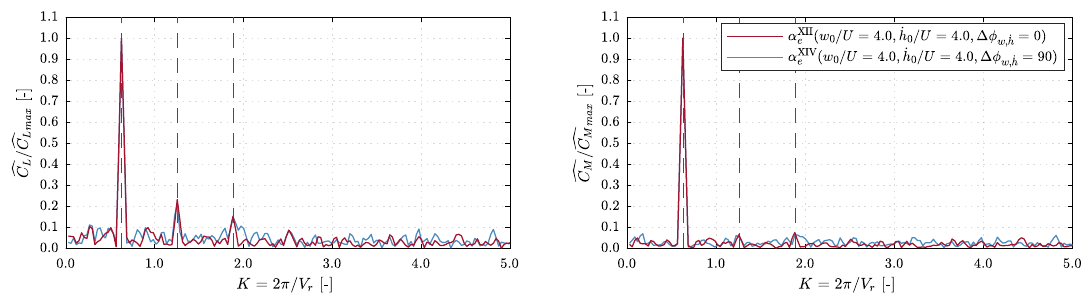} 
\vspace*{-0.6cm}
\caption{Bluff deck: Normalized FFT of the lift $C_L$ (left) and moment $C_M$ (right)  coefficients for the CFD  model of the combined forcing cases at $V_r$=10.  The broken vertical lines denote the main, the second, and the third harmonics $K$ values.  Angle values are in [deg]. (For interpretation of the references to color in this figure legend, the reader is referred to the web version of this article.) }
\label{spectralPlot_vr_10}
\end{figure}
\FloatBarrier 
In the case of the LU model, the separate components are also depicted, which were determined to be individually linear in the previous section. As seen from the figure, there are discrepancies between the CFD and the LU model particularly for the lift force. This effect is further manifested in the FFT for both cases (XII and XIV: zero and 90 deg phase) in Fig.~\ref{spectralPlot_vr_10}, where a difference in the main harmonic between the models can be noted with additional higher-order components for the CFD model that indicate aerodynamic nonlinearities.\par

\begin{figure}[h!]
\centering
\includegraphics[trim=0cm 0.3cm 0cm 0cm,clip,width=1\columnwidth]{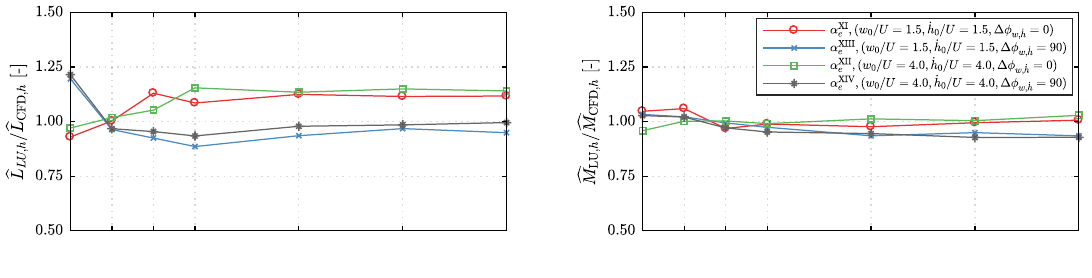} 
\includegraphics[trim=0cm 0cm 0cm 0cm,clip,width=1\columnwidth]{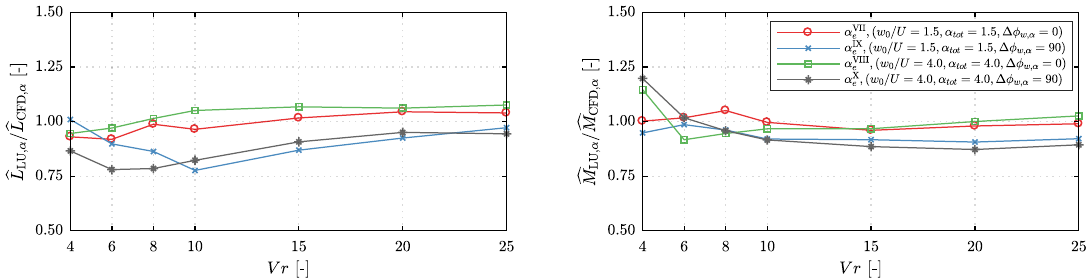} 
\caption{Streamlined deck: Spectral amplitude ratio of the main harmonic between the LU and CFD models for combined input: Gust and heave (top); gust and pitch (bottom). Angle values are in [deg]. }
\label{GB_PS_ampRatio}
\end{figure}

\begin{figure}[h!]
\centering
\includegraphics[trim=0cm 0.3cm 0cm 0cm,clip,width=1\columnwidth]{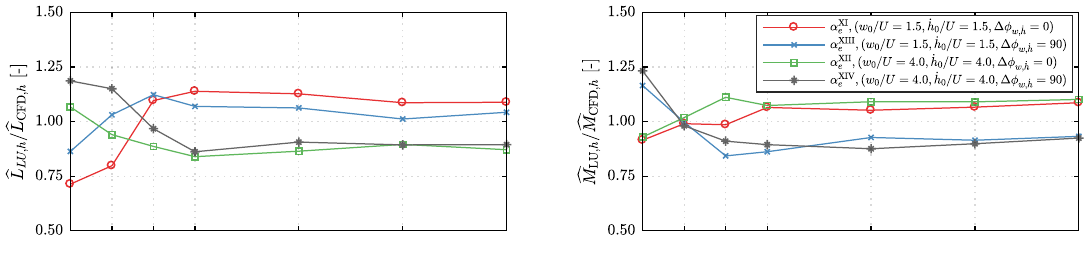} 
\includegraphics[trim=0cm 0cm 0cm 0cm,clip,width=1\columnwidth]{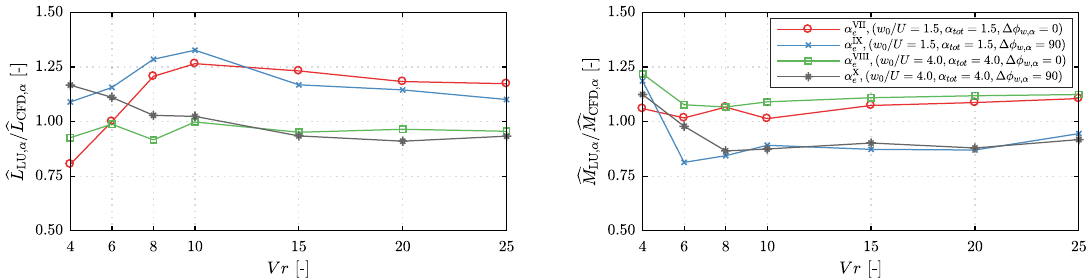} 
\caption{Bluff deck: Spectral amplitude ratio of the main harmonic between the LU and CFD models for combined input: Gust and heave (top); gust and pitch (bottom). Angle values are in [deg].}
\label{MG_PS_ampRatio}
\end{figure}
\FloatBarrier
The main harmonic spectral ratios between the LU and CFD models are shown in Fig.~\ref{GB_PS_ampRatio} for the streamlined and in Fig.~\ref{MG_PS_ampRatio} for the bluff deck. The moment force abides the linear hypothesis for the streamlined deck (see Fig.~\ref{GB_PS_ampRatio}, left), while differences up to 25 \% can be noted for the moment of the bluff deck at low reduced velocities. Interestingly, the effect of the phase between the motion and gust, rather than the amplitudes of these individual actions, is more detrimental for this effect. The effect in the lift is more severe, especially for the bluff deck, where differences in the main harmonic of more than 25\% can be observed. Unlike for the moment, here it is the amplitude of the total angle that impacts the differences that occur mainly at low reduced velocities $V_r$. As discussed in the next section, the high-frequency deck oscillation (i.e. low reduced velocities) causes the shedding of vortices from the leading and trailing edges at a higher rate than the wake flow convection. This, in turn, impacts the output aerodynamic forces causing an increase and time lag in the instantaneous pressure, which can be as nonlinear effects.\par

\begin{figure}[h!]
	\centering
	\includegraphics[trim=0cm 0cm 0.8cm 0cm,clip,width=0.88\columnwidth]{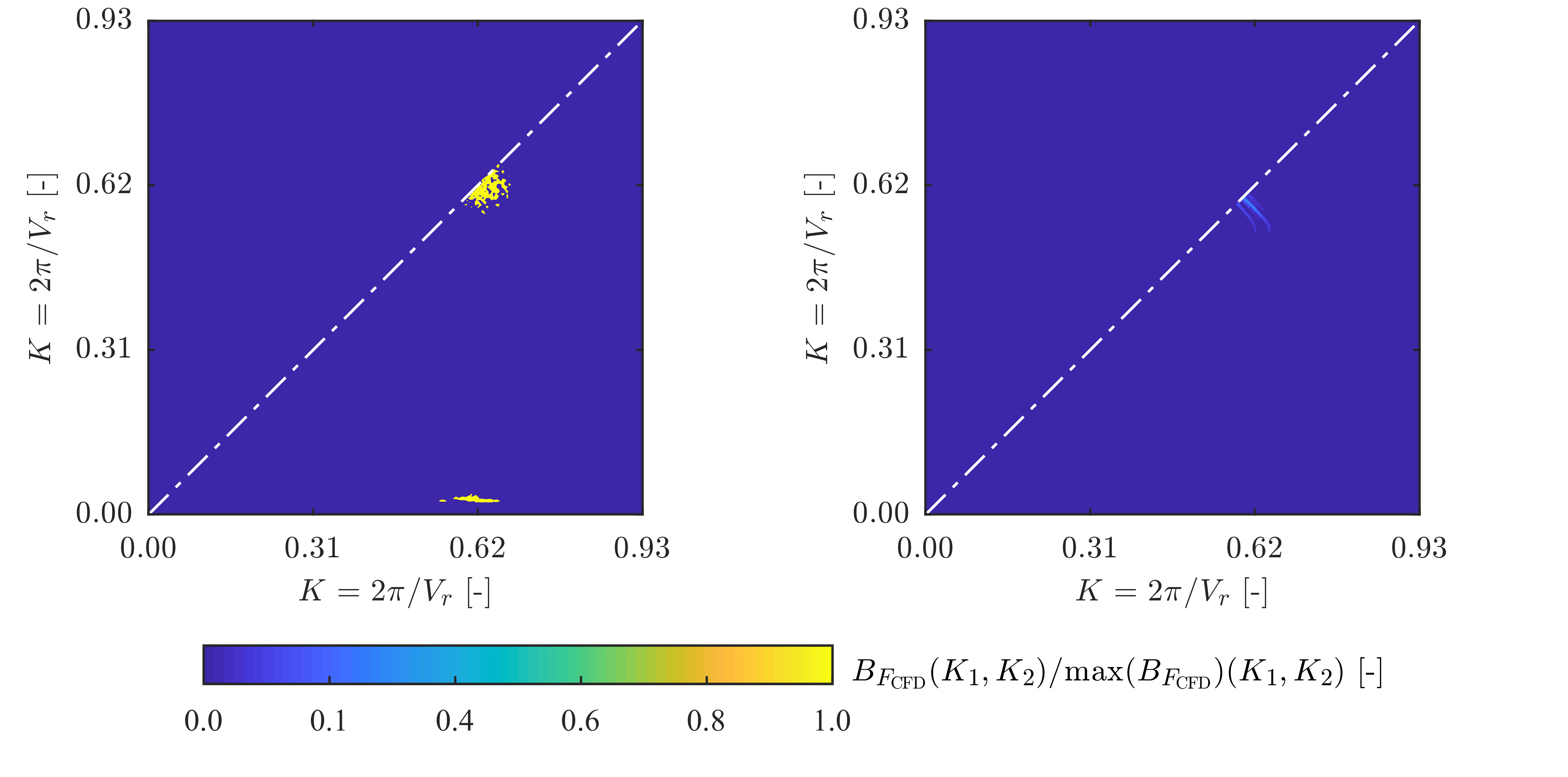}
	\vspace*{-0.23cm} 
	\caption{Bluff deck: Normalized bispectrum amplitude of the CFD lift $C_L$ (left) and moment $C_M$ (right) coefficients for combined forcing case $\alpha_{e}^{\mathrm{XII}}$ ($\dot{h}_0/U=4,\; \mathrm{deg}, w_0/U=4 \; \mathrm{deg}, \Delta \phi_{w,\dot{h}} = 0 \; \mathrm{deg}$) at $V_r$ =10.  (For interpretation of the references to color in this figure legend, the reader is referred to the web version of this article.) }
	\label{BS_plot}
\end{figure}

Apart from impacting the first harmonic,  the nonlinear interaction between the gust- and motion-induced forces yields higher-order harmonics in the case of the bluff deck section (see Tab.~\ref{T2}). This behavior is dependent on the reduced velocity $V_r$. For $V_r$=4 and 25, both the lift and moment force depict nonlinear behavior, generally the moment being stronger for $V_r$=4 and the lift for $V_r$=25. An interesting feature is portrayed in the moment at $V_r$=4 for the cases VIII ($w_0/U=4\; \mathrm{deg}, \dot{h}_0/U=4 \; \mathrm{deg}, \Delta \phi_{w,\dot{h}}=0\; \mathrm{deg}$) and X ($w_0/U=4\; \mathrm{deg}, \dot{h}_0/U=4 \; \mathrm{deg}, \Delta \phi_{w,\dot{h}}=90\; \mathrm{deg}$), where it is the phase-shift between the motion and the gust that increases the contribution of the higher order harmonics from 16.4\% to 35.6\%. This indicates the nonlinearity may not originate only based on the angle of attack, rather than the complicated flow features and breaking of shear layer as discussed in the following section. The higher-order harmonics at $V_r$=10 are more prominent in the lift forces and can up go to 23\%  of the main (forcing) harmonic for the bluff deck. To ensure that the second-order are due to nonlinear interaction, Fig.~\ref{BS_plot} depicts the bispectrum of the lift and moment of the CFD model for the XII case ($w_0/U=4\; \mathrm{deg}, \dot{h}_0/U=4 \; \mathrm{deg}, \Delta \phi_{w,\dot{h}}=0\; \mathrm{deg}$). A clear concentration of spectral density can be observed at the forcing reduced frequency $K\approx0.63$ (corresponding to $V_r$=10). As the phased-filtered bispectrum is presented~\citep{Kavrakov2020}, it means that there is zero phase between the main and second-order harmonic, which is a nessesary condition for nonlinearity~\citep{Fackrell1995}. Similar indications were noted bispectra of the other cases of combined effects due to gust and motion (omitted for the sake of brevity).\par
\begin{table}[h!]
\centering
\includegraphics[trim=0.25cm 9cm 0.1cm 9.1cm,clip,width=1\columnwidth]{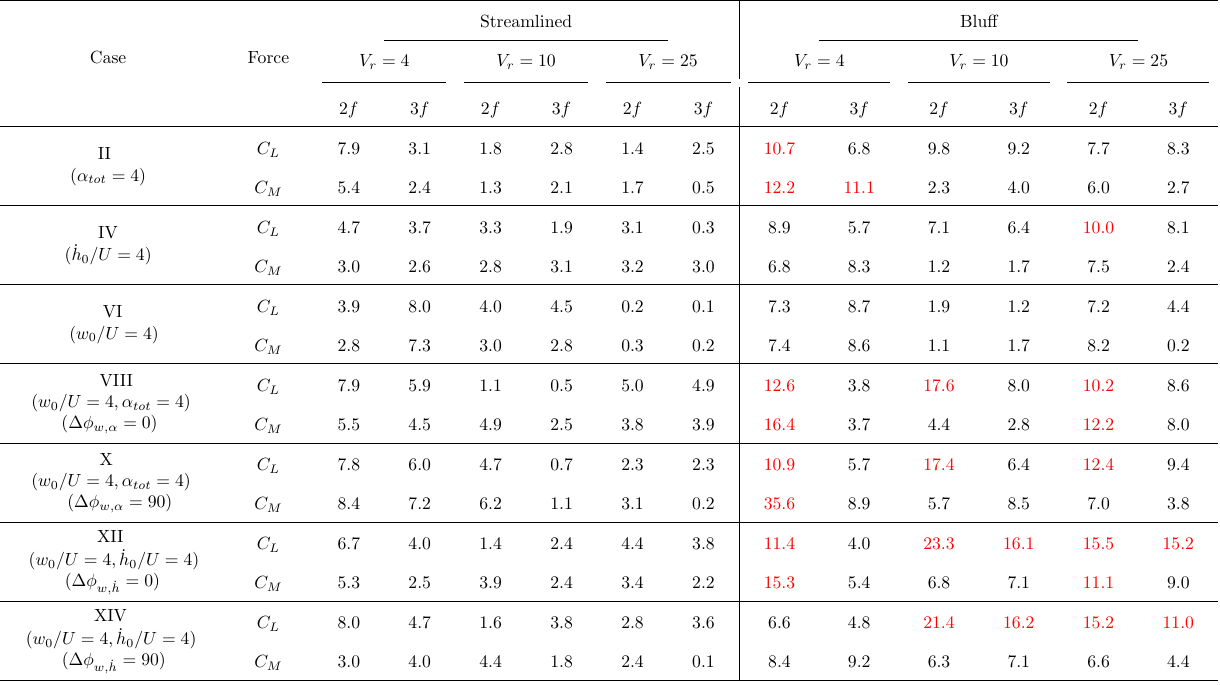} 
\caption{ FFT amplitude ratio between the main harmonic with frequency $f$ and higher-order harmonics of 2$f$ and 3$f$ of the lift $C_L$ and moment $C_M$ coefficients of selected cases. The ratio is in [\%], while the angle values are in [deg].  A threshold of 10\% is highlighted in red to distinguish between nonlinearity and spectral noise.}
\label{T2}
\end{table}

\begin{figure} [h!]
\includegraphics[trim=0cm 17.5cm 0cm 2.8cm,clip,width=0.88\columnwidth]{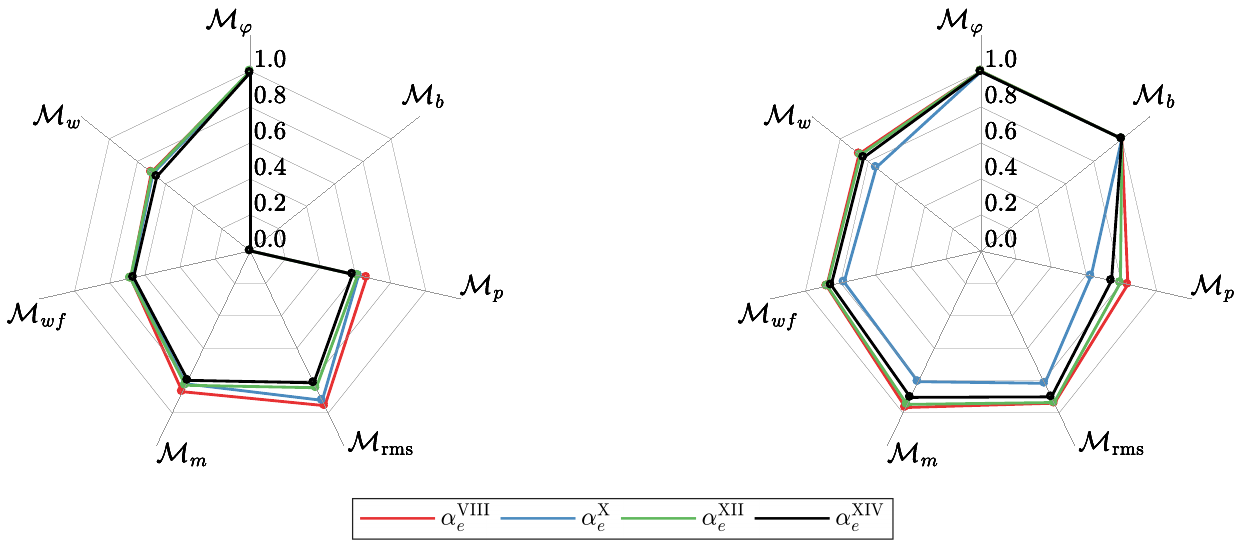} 
    \caption{Bluff deck: Comparison metrics between the time-histories of the LU and CFD models (CFD is reference) for combined action (gust and motion) $C_L$ (left) and $C_M$ (right) at $V_r$ =10 for the following cases: $\alpha_{e}^{\mathrm{VIII}}$ ($\alpha_{tot}=4 \; \mathrm{deg}, w_0/U=4 \; \mathrm{deg}, \Delta \phi_{w,\alpha} = 0 \; \mathrm{deg}$), $\alpha_{e}^{\mathrm{X}}$ ($\alpha_{tot}=4 \; \mathrm{deg}, w_0/U=4 \; \mathrm{deg}, \Delta \phi_{w,\dot{h}} = 90 \; \mathrm{deg}$), $\alpha_{e}^{\mathrm{XII}}$ ($\dot{h}_0/U=4 \; \mathrm{deg}, w_0/U=4 \; \mathrm{deg}, \Delta \phi_{w,\dot{h}} = 0 \; \mathrm{deg}$) and $\alpha_{e}^{\mathrm{XIV}}$ ($\dot{h}_0/U=4 \; \mathrm{deg}, w_0/U=4 \; \mathrm{deg}, \Delta \phi_{w,\dot{h}} = 90 \; \mathrm{deg}$). (For interpretation of the references to color in this figure legend, the reader is referred to the web version of this article.) }
    \label{metricsPlot}
\end{figure}

The discrepancies in the main and higher-order harmonics are a clear indication of the effect of nonlinear interaction in the forces due to gust and motion. To quantify this effect further, the comparison metrics of the aerodynamic forces for the CFD and LU models of the bluff deck are presented in Fig.~\ref{metricsPlot} for the cases due to combined gust and motion excitation at high amplitudes (VII, X, XII, and XIV) at $V_r$=10, taking the CFD model as a reference.
Most of the metrics for the moment force are above $\mathcal{M}\geq0.8$, which can be considered fair a correspondence and that the linear hypothesis is generally valid (\cite{Kavrakov2020} gives approximate threshold of 0.9 as a good agreement). However, in case of the lift force, the wavelet $\mathcal{M}_w$ and $\mathcal{M}_wf$ and the peak metric $\mathcal{M}_p$ are below the 0.8 threshold. Additionally, the wavelet metrics are different between each other indicated magnitude as well as frequency modulation. As expected, the bispectrum metrics are zero $\mathcal{M}_b=0$, since the LU model cannot capture second-order harmonics.\par
The investigation in this section showed that the nonlinear interaction between the gust- and motion-induced effects can influence the aerodynamic forces mostly in case of bluff decks.

\FloatBarrier

\subsection{Influence of free-stream gusts on the shear layer of a moving deck}
\label{Sec7_Press}
\begin{figure}[h!]
  \centering
  \includegraphics[width=1\linewidth]{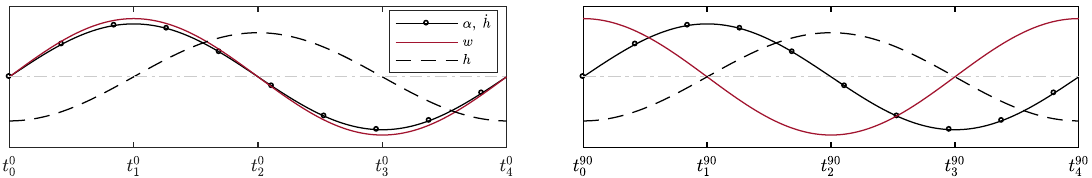}
\caption{Schematic of the input wind and motion time-histories during one cycle, based on the sign convention depicted in Fig.~\ref{Fig2}: $\Delta \phi_{w, j } = 0$  deg left and $\Delta \phi_{w, j }  = 90 $ deg  right,  where $j \in $ \{$\dot{h}, \alpha$\}.}
 \label{TimeInstant}
\end{figure}

\begin{figure}[h!]
%\vspace*{-1cm}
\centering
  \includegraphics[trim=0.21cm 10.8cm 0.5cm 7.3cm,clip,width=1.01\textwidth,right]{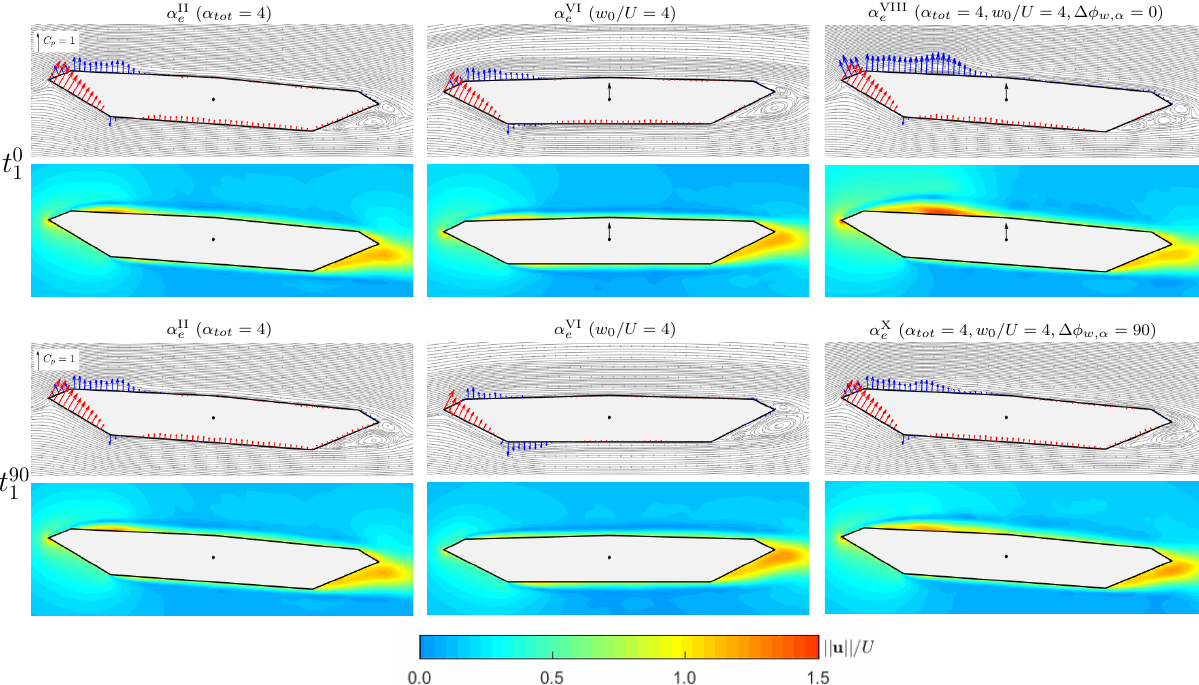}
\caption{Streamlined deck: Instantaneous pressure and velocity field averaged over 32 cycles (see Fig.~\ref{TimeInstant}) for forcing at $V_r=10$. The section is forced with pitch motion (left, $\alpha_{e}^{\mathrm{II}}$ with $\alpha_0 =$ 3.5 deg), gust (center, $\alpha_{e}^{\mathrm{VI}}$ with $\alpha_0 =$ 3.5 deg), and combined pitch and gust with zero phase (top right, $\alpha_{e}^{\mathrm{VIII}}$ with $\alpha_0 =$ 3.5 deg) and phase-distorted (bottom right, $\alpha_{e}^{\mathrm{X}}$ with $\alpha_0 =$ 3.5 deg). The black arrow in the center of the section represents the direction of the gust at the selected time time instant. Angle values are in [deg]. (For interpretation of the references to color in this figure legend, the reader is referred to the web version of this article.) }
 \label{GB_P_Pr}
\end{figure}

\begin{figure}[h!]

%\flushright
 \centering
  \includegraphics[trim=0.21cm 14.45cm 0.5cm 3.55cm,clip,width=1.00\textwidth,right]{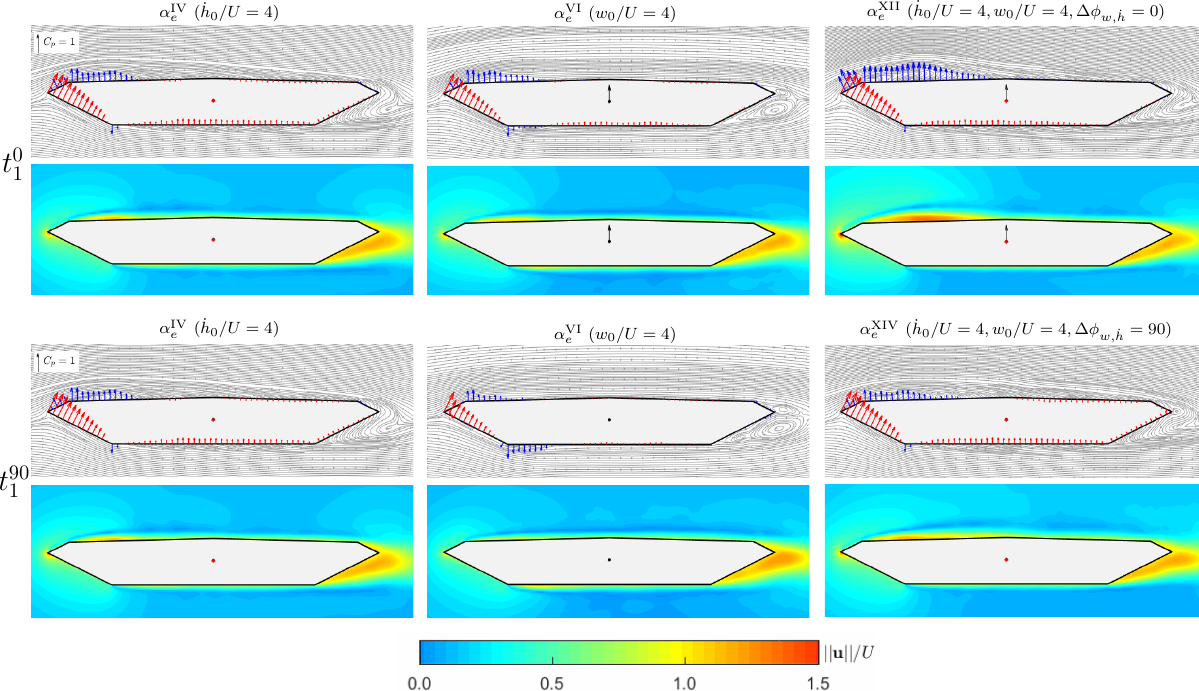}
 \caption{Streamlined deck: Instantaneous pressure and velocity field averaged over 32 cycles (see Fig.~\ref{TimeInstant}) for forcing at $V_r=10$. The section is forced with heave motion  (left, $\alpha_{e}^{\mathrm{IV}}$), gust (center, $\alpha_{e}^{\mathrm{VI}}$), and combined heave and gust with zero phase (top right, $\alpha_{e}^{\mathrm{XII}}$) and phase-distorted (bottom right, $\alpha_{e}^{\mathrm{XIV}}$). The black arrow in the center of the section represents the direction of the gust at the selected time time instant. Angle values are in [deg]. (For interpretation of the references to color in this figure legend, the reader is referred to the web version of this article.) }
  \label{GB_H_Pr}
\end{figure}

\begin{figure}[h!]
%\vspace*{-0.5cm}
%\flushright
   \includegraphics[trim=0.1cm 16.03cm 0.5cm 3.55cm,clip,width=1.01\textwidth,right]{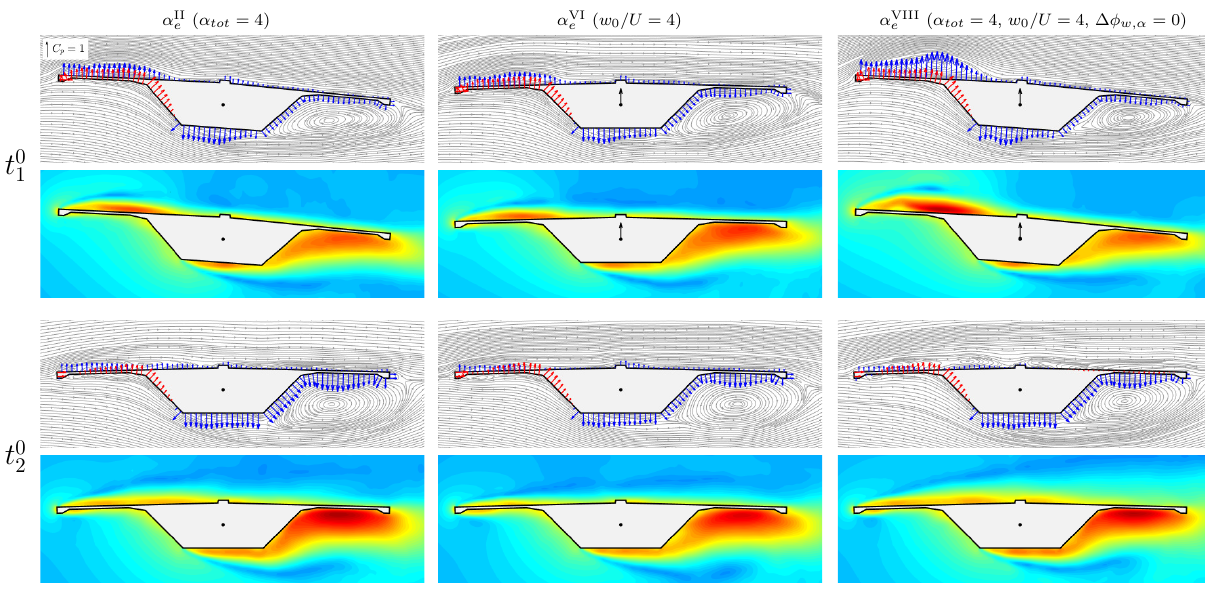}
 
\hbox{ \hspace{5.2cm}
\centering
 \includegraphics[trim=0cm 0cm 0cm 0cm,clip,width=0.5\columnwidth]{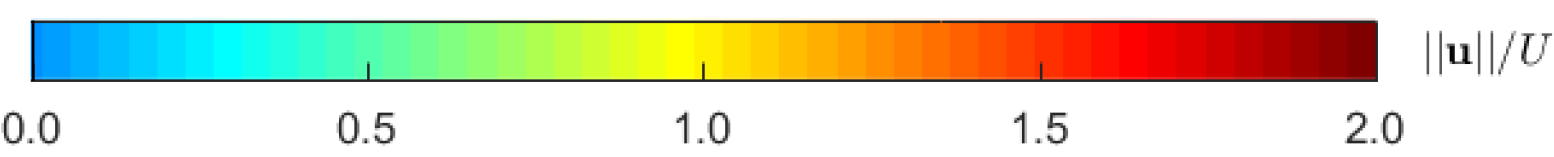}
        } 
\caption{Bluff deck: Instantaneous pressure and velocity field averaged over 32 cycles (see Fig.~\ref{TimeInstant}) for forcing at $V_r=10$. The section is forced with pitch motion (left, $\alpha_{e}^{\mathrm{II}}$ with $\alpha_0 =$ 3.5 deg), gust (center, $\alpha_{e}^{\mathrm{VI}}$), and combined pitch and gust with zero phase (right, $\alpha_{e}^{\mathrm{VIII}}$ with $\alpha_0 =$ 3.5 deg). The black arrow in the center of the section represents the direction of the gust at the selected time time instant.  Angle values are in [deg]. (For interpretation of the references to color in this figure legend, the reader is referred to the web version of this article.) }
 \label{MG_P_Pr_1}
\end{figure}

Figure \ref{GB_P_Pr} to \ref{MG_H_Pr_2} show the pressure and velocity fields for both decks at the time instants (see Fig.~\ref{TimeInstant}), where the leading edge flow separation and the subsequent reattachment is prominent. The pressure and velocity field plots represent a mean value of 32 time instants where the bridge deck (or gust) reaches the same state/position during each successive cycle. For example,  in Fig.~\ref{GB_P_Pr}, the top left plot $t_1^0$ represents an average pressure and velocity field of  32 time instants at which the bridge deck is at a maximum nose-up position, as shown in Fig.~\ref{TimeInstant}. The same applies to the gust-induced  setup (see Fig.~\ref{GB_P_Pr} top center): the figure represents an average of 32 time instants at which the gust is at the maximum positive state (see Fig.~\ref{TimeInstant}). The ﬂuctuating pressure amplitudes are normalized to the stagnation pressure, and $||$\textbf{u}$||$ is the mean fuctuating velocity magnitude. 

%\vspace*{-1.0cm}
\par 
The instantaneous pressures and velocity fields for the streamlined deck for the in-phase case ($\Delta \phi_{w, j }  = 0$ deg, Figs. \ref{GB_P_Pr} and \ref{GB_H_Pr}), show that the gust moderately stretches the leading edge shear layer towards the windward direction and shifts the peak pressure location to the reattachment points. On the other hand, the out-of-phase case ($\Delta \phi_{w, j }  = 90$ deg) dissolves the leading edge separation vortices, causing earlier reattachment. The latter is in line with the findings \citep{HaanJr2009}, who reported that the small-scale turbulence shortens the separation layer. Evaluating the combined excitation setup pressure distributions for the two cases ($\Delta \phi_{w, j }  = 0$ deg and $\Delta \phi_{w, j }  = 90$ deg) shows the phase shift effect is primarily localized on the top chord of the deck. Regardless of the DOF for the motion, the bottom chord combined excitation pressure distribution is roughly the same for both phase shift cases. As the lift force is produced as the sum of the upper and lower surface pressures of the deck, the phase shift changes the lift force more than the moment force, as shown in the spectral amplitude ratio (see Fig.~\ref{GB_PS_ampRatio}). The shortening of the lever arm alleviates the phase shift effect on the moment force. However, in a qualitative sense, the linear assumption holds for the streamlined deck as the velocity and pressure fields for the combined excitation cases appear to be a superposition of the corresponding fields for the individual excitation cases.
%\FloatBarrier
\begin{figure}[h!]
 
\includegraphics[trim=0.1cm 5.8cm 0.3cm 13.5cm,clip,width=1.01\textwidth,right]{Fig30}

\hbox{ \hspace{5.2cm}
\centering
 \includegraphics[trim=0cm 0cm 0cm 0cm,clip,width=0.5\columnwidth]{Fig32}
        } 

\caption{Bluff deck: Instantaneous pressure and velocity field averaged over 32 cycles (see Fig.~\ref{TimeInstant}) for forcing at $V_r=10$. The section is forced with pitch motion (left, $\alpha_{e}^{\mathrm{II}}$ with $\alpha_0 =$ 3.5 deg), gust (center, $\alpha_{e}^{\mathrm{VI}}$), and combined phase-distorted pitch and gust (right, $\alpha_{e}^{\mathrm{X}}$ with $\alpha_0 =$ 3.5 deg). The black arrow in the center of the section represents the direction of the gust at the selected time time instant. Angle values are in [deg]. (For interpretation of the references to color in this figure legend, the reader is referred to the web version of this article.) }
 \label{MG_P_Pr_2}
\end{figure}

\begin{figure}[h!]

   \includegraphics[trim=0.6cm 16.cm 0.3cm 4cm,clip,width=1.01\textwidth,right]{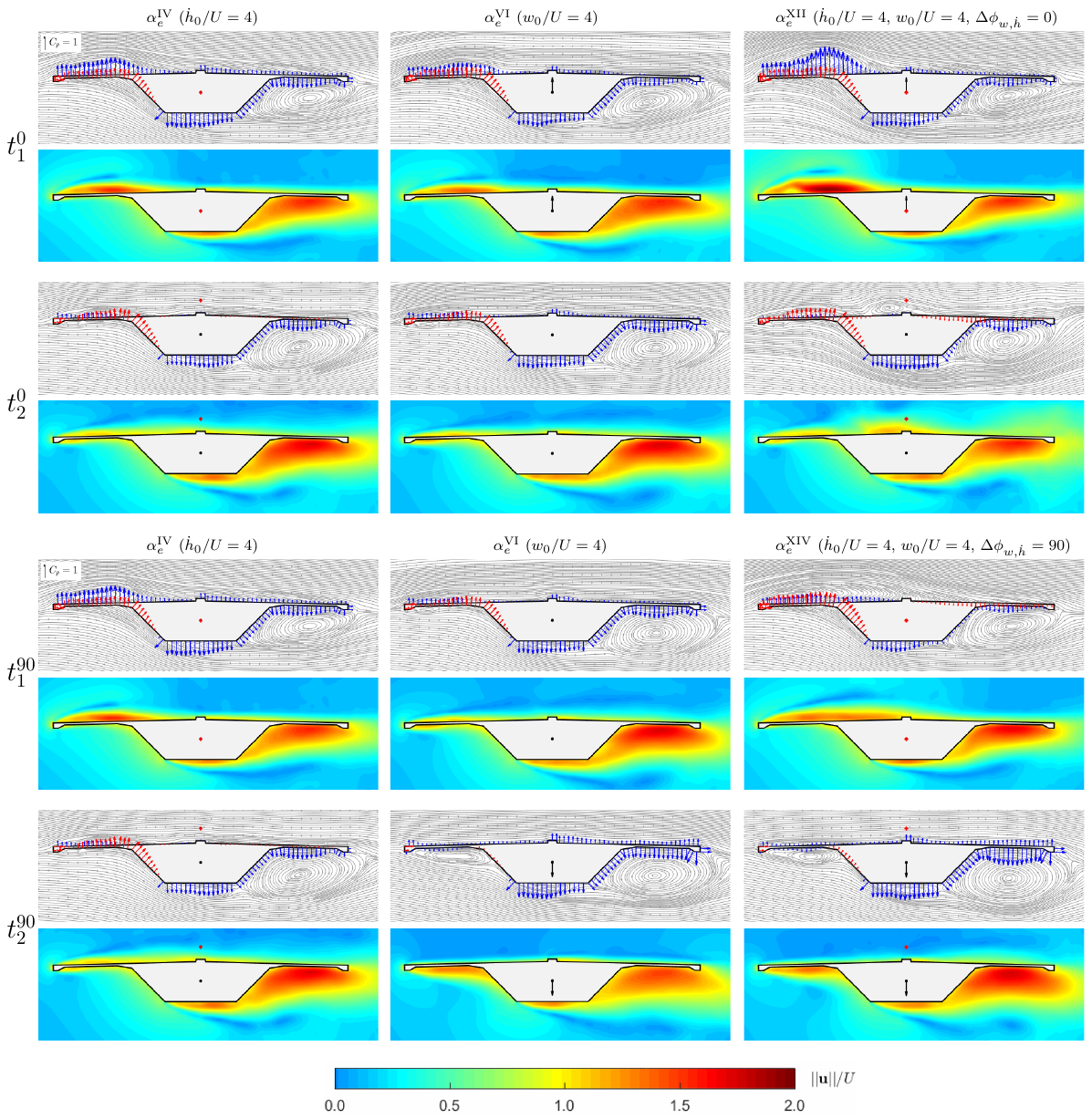}
  
         \hbox{ \hspace{5.2cm}
\centering
 \includegraphics[trim=0cm 0cm 0cm 0cm,clip,width=0.5\columnwidth]{Fig32}
         } 
\caption{Bluff deck: Instantaneous pressure and velocity field averaged over 32 cycles (see Fig.~\ref{TimeInstant}) for forcing at $V_r=10$. The section is forced with heave motion (left, $\alpha_{e}^{\mathrm{IV}}$), gust (center, $\alpha_{e}^{\mathrm{VI}}$), and combined heave and gust with zero phase (right, $\alpha_{e}^{\mathrm{XII}}$). The black arrow in the center of the section represents the direction of the gust at the selected time time instant. Angle values are in [deg]. (For interpretation of the references to color in this figure legend, the reader is referred to the web version of this article.) }
 \label{MG_H_Pr_1}
\end{figure}
\FloatBarrier
\begin{figure}[h!]

\centering

   \includegraphics[trim=0.6cm 6.2cm 0.3cm 13.5cm,clip,width=1.01\textwidth,right]{Fig31}

\hbox{ \hspace{5.2cm}
\centering
 \includegraphics[trim=0cm 0cm 0cm 0cm,clip,width=0.5\columnwidth]{Fig32}
        } 
         
\caption{Bluff deck: Instantaneous pressure and velocity field averaged over 32 cycles (see Fig.~\ref{TimeInstant}) for forcing at $V_r=10$. The section is forced with heave motion (left, $\alpha_{e}^{\mathrm{IV}}$), gust (center, $\alpha_{e}^{\mathrm{VI}}$), and combined phase-distorted heave and gust (right, $\alpha_{e}^{\mathrm{XIV}}$). The black arrow in the center of the section represents the direction of the gust at the selected time time instant. Angle values are in [deg]. (For interpretation of the references to color in this figure legend, the reader is referred to the web version of this article.) }
 \label{MG_H_Pr_2}
\end{figure}

\begin{figure}[h!]
	\centering
	  \includegraphics[width=0.8\linewidth]{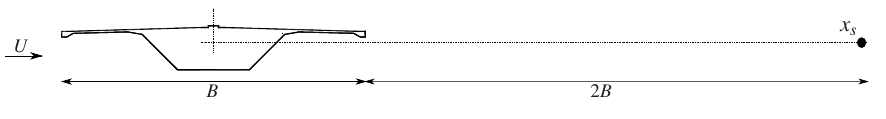}
	\includegraphics[width=1\linewidth]{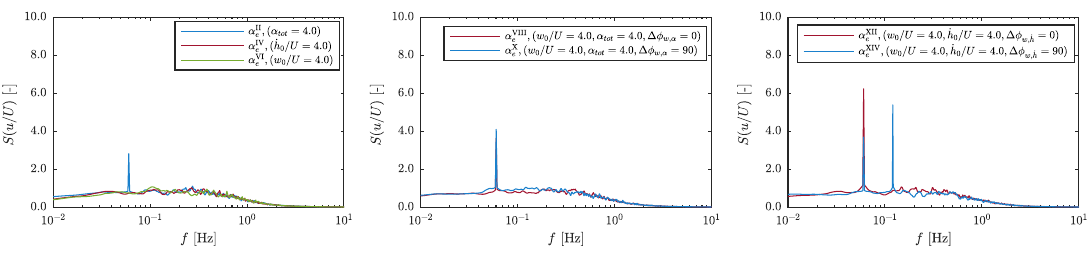}
	\includegraphics[width=1\linewidth]{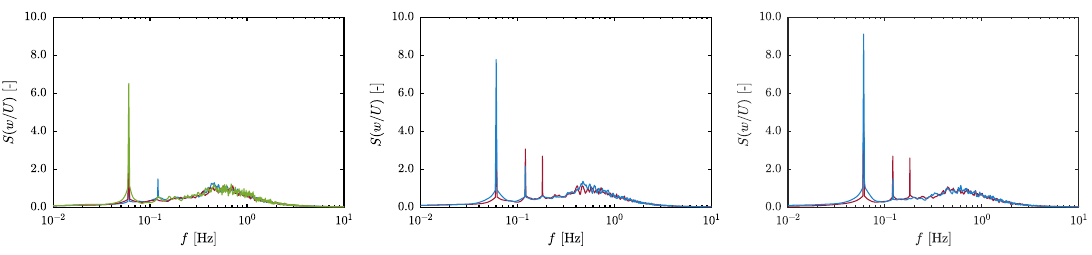}
	\caption{Bluff deck:  Power spectral density of the downwash velocity at  sampling location $x_s$ (top) for forcing at $V_r$=10: longitudinal component $u$ (middle) and vertical component $w$ (bottom). Angle values are in [deg]. (For interpretation of the references to color in this figure legend, the reader is referred to the web version of this article.) 
	}
	\label{fft_sheding}
\end{figure}
%\FloatBarrier
 The shear layer of the bluff deck (see Fig.~\ref{MG_P_Pr_1} to \ref{MG_H_Pr_2}) is influenced by the unsteady motion- gust- induced vortices generated at the leading edge and a strong Kármán vortex type bubble at the bottom trailing end. In the case of combined action of motion and gust with $\Delta \phi_{w,j} = 0$ deg for motion in both DOFs (see Fig.~\ref{MG_P_Pr_1} and \ref{MG_H_Pr_1}), the flow experiences a strong separation at the leading edge that is then followed by abrupt reattachment. This is not the case when only one individual component (gust or motion) is acting. Moreover, multiple reattachment points can be observed in some instances when the combined effect of gust and motion (both gust and motion are acting) (see Fig.~\ref{MG_H_Pr_1} at $t_{2}^{0}$). This may explain the excitation of the higher-order frequencies noted in the lift forces. Over the bottom surface of the deck, where the sharp edge separation of flow is prominent, the pressure distribution depends on the separation vortex in particular, and the gust has a minor impact. 

% \begin{figure}[h!]
%  \centering
%  \includegraphics[width=0.8\linewidth]{Fig33}
%\caption{Sampling location of the downwash velocity used in Figs.~\ref{fft_sheding} and~\ref{VXShedngPic}. }
% \label{VXShedngSectn}
%\end{figure}
\par
\begin{figure}[h!]
  \centering

  \includegraphics [trim=0.1cm 12cm 0.1cm 11.5cm,clip,width=1.01\textwidth,right]{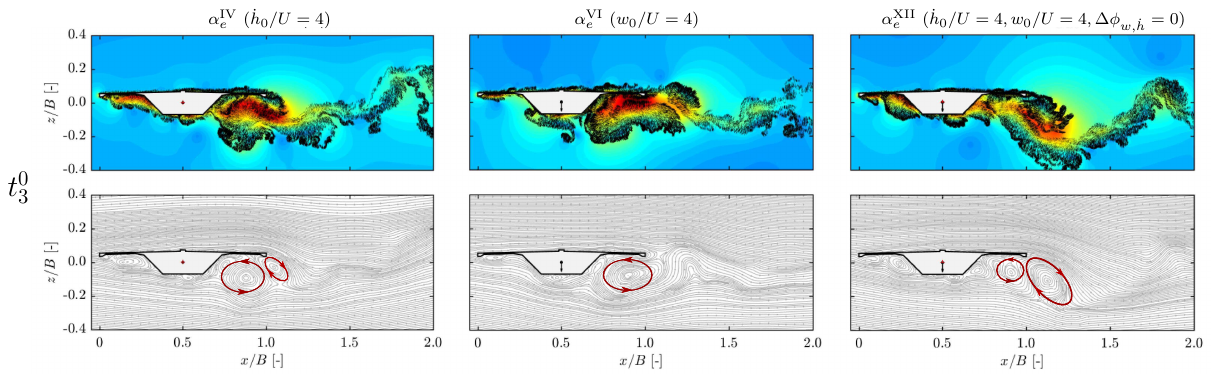} 
\caption{Bluff deck: Instantaneous particle map with velocity field (top) and streamlines (bottom) for forcing at $V_r$=10. The section is forced with heave motion (left, $\alpha_{e}^{\mathrm{IV}}$), gust (center, $\alpha_{e}^{\mathrm{VI}}$), and combined heave and gust with zero phase (right, $\alpha_{e}^{\mathrm{XII}}$). The black arrow in the centre of the section represent the direction of the gust at that time instant  while the red-cross mark (\textcolor{red}{$+$}) shows the position of the deck stiffness centre at stationary position.  Angle values are in [deg]. }
 \label{VXShedngPic}
\end{figure}

Further, the influence of the large-scale gusts acting on moving bodies is manifested in the wake flow. Figure~\ref{fft_sheding} shows the spectrum of the filtered downwash velocities $u$ and $w$ recorded at 2$B$ from the trailing edge. In the case of combined action of gust and motion, nonlinear interaction in the wake can be observed through the higher-order harmonics present in the $u$ and $w$ velocity components. This multi-frequency shedding originates from the interaction of the impinging leading and trailing-edge separation vortices \citep{Deniz1998, McRobie2013, Matsumoto2008}.  Hence, the gust modifies both the rate of shedding and the strength of motion-induced vortices. One such instant is presented in Fig.~\ref{VXShedngPic}, where the gust enhances the entrainment process causing strong shedding from the bottom edge. In the case of the separate effect of gust or motion, a considerable portion of the bubble remains intact. From the above discussion, it can be concluded that the gust can have a considerable effect on the shear-layer flow dynamics, depending on the deck type. 
%\FloatBarrier

\section{Conclusion}
\label{Sec7_Conlusion}

The study systematically investigated the nonlinear interaction, or the lack thereof, between the motion- and gust-induced forces acting on bluff bodies. A CFD methodology based on the VPM was devised to simulate such nonlinear interaction by subjecting a sinusoidally oscillating bluff body to a large scale sinusoidal gust with identical frequency. Initially, the methodology was applied to a flat plate. By comparing the aerodynamic forces of the CFD model with their linear analytical counterpart, it was shown that the linear superposition principle holds, thereby verifying the proposed methodology. Next, the methodology was applied to streamlined and bluff bridge decks to excite the nonlinear interaction. Two aspects were considered: the nonlinear dependence of the aerodynamic forces on the combined angle of attack due to motion and gust and the influence of the gusts on the shear layer on a moving body. The LU model was used as a baseline for the linear aerodynamic forces.\par
The study of the aerodynamic forces for the streamlined deck indicated that the moment was generally linear, while discrepancies up to 25\% in the main harmonic were recorded for the lift with respect to the linear baseline. Nevertheless, no high-order force harmonics were noted of the streamlined deck. On the contrary, for the bluff deck, both lift and moment exhibited nonlinear behavior, which manifested itself in two manners. First, discrepancies in the main harmonic with respect to the linear baseline were noted up to 25-30\%. Second, higher-order harmonics emerged, resulting in values up to 35\% with respect to the main forcing harmonic. The phase filtered bispectrum was used to verify that the second-order harmonics were a nonlinear feature rather than spectral noise. It was observed that the nonlinear features were not purely dependent on the amplitude of the effective angle of attack but also on the phase between the motion and the gust. \par
Qualitative observation of the mean instantaneous pressure and velocity ﬁeld revealed large-scale vertical gust slightly stretches the leading edge separation length when the input signals are in phase.  The phase-distorted input signals dissolve the leading edge separation vortices, causing earlier reattachment. The streamlined deck combined gust and motion action results in slightly stronger leading-edge separation of flow as compared with the individual (gust or motion) excitations.  However, in a general sense, the velocity and pressure fields for the combined excitation cases appear to be a superposition of the corresponding fields for the individual excitation cases. In the case of the bluff deck, the effect of the large-scale gust on the shear layer varies on the locale of the deck surface. On the top chord, the combined action of gust and motion induces strong leading-edge flow separation, which is then followed by abrupt reattachment.  Compared to the combined pitch and gust action, the heave and gust combination exhibits strong, in some instants multiple-point, reattachment.  In cases of individual excitation (gust or motion), the leading edge separation is relatively small and does not exhibit prominent reattachment.  Over the bottom trailing surface of the deck, the flow separates from the sharp edges resulting in a strong air bubble trapped between the shear layer and the deck surface. In this area, the pressure distribution depends on the separation vortex in particular and the free-stream gust has a minor impact on the shear layer. Nonetheless, downwash velocity recordings indicate that the synergistic interaction of free-stream gust and motion modifies the strength and shedding rate of the separation vortices.  This is evident from the higher amplitude, multi-frequency downwash velocity recordings of the combined gust and motion cases, compared to individual gust or motion excitations. These intricate flow aspects of the bluff deck could explain the nonlinearity observed in the forces of the combined gust- and motion-induced cases.  
\par
In conclusion, the results of this study show that the nonlinear interaction between the gust- and motion-induced forces can be prominent for bluff bodies with sharp edges that cause strong separation. This nonlinear interaction originates from the complex shear-layer flow dynamics, rather than purely on the amplitude of the effective angle of attack, and is dependent on the gust and motion forcing frequency. The recorded nonlinear behavior of the bluff section necessitates the need for nonlinear analytical predictive models. Although the study is purely numerical based on CFD, it aims to provide a deeper understanding of the complex fluid-interaction phenomena occurring when moving bodies are subjected to free-stream turbulence. It is intended that its outcome will serve to initiate further experimental studies, which are warranted to substantiate the presented results.

\section*{Acknowledgements}
ST gratefully acknowledges the support part by the Bauhaus Research School and the Smart Neighbourhood (smood) initiative. IK and GM gratefully acknowledge the support by the German Research Foundation (DFG) [Project No. 329120866].  

\bibliographystyle{model2-names}
\biboptions{comma,numbers,sort&compress}
\bibliography{library}

\end{document}